\documentclass[12pt,eqsecnum,preprint,flushrt]{aastex}
\slugcomment{Submitted to ApJ}
\usepackage{amssymb}
\begin{document}
\title{Mean-Field Magnetohydrodynamics of Accretion Disks}
\author{Frank H. Shu$^1$, Daniele Galli$^2$, Susana Lizano$^3$, Alfred E. 
Glassgold$^4$, Patrick H. Diamond$^1$}
\affil{$^1$Department of Physics, University of California, San Diego, CA 
92093\\
$^2$INAF-Osservatorio Astrofisico di Arcetri, 
Largo E. Fermi 5, Firenze I-50125, Italy\\
$^3$CRyA, Universidad Nacional Aut\'onoma de M\'exico, 
Apdo. Postal 72-3, 58089 Morelia, Mexico \\
$^4$Astronomy Department, University of California, Berkeley, CA 94720}
\email{fshu@physics.ucsd.edu}

\begin{abstract}
We consider the accretion process in a disk with magnetic fields 
that are dragged in from the interstellar medium by gravitational collapse. 
Two diffusive processes are at work in the system: 
(1) ``viscous" torques exerted by turbulent and magnetic stresses, and (2) 
``resistive" redistribution of mass with respect to the magnetic flux arising 
from the imperfect conduction of current. In steady state, self-consistency 
between the two rates of drift requires that a relationship exists between the 
coefficients of turbulent viscosity and turbulent resistivity.  Ignoring any interactions
with a stellar magnetosphere, we solve the steady-state equations for a magnetized
disk under the gravitational attraction of a mass point and threaded by an
amount of magnetic flux consistent with calculations of magnetized gravitational
collapse in star formation. Our model mean-field equations have an exact analytical
solution that corresponds to magnetically diluted Keplerian rotation about the central mass
point.  The solution yields the strength of the magnetic field and the surface density as
functions of radial position in the disk and their connection with the departure from pure
Keplerian rotation in representative cases.  We compare the predictions of the theory with
the available observations concerning T Tauri stars, FU Orionis stars, and low- and high-mass
protostars. Finally, we speculate on the physical causes for high and low states of the accretion
disks that surround young stellar objects.  One of the more important results of this study is the
physical derivation of analytic expressions for the turbulent viscosity and turbulent resistivity.

\end{abstract}

\newcommand{\be}{\begin{equation}}
\newcommand{\ee}{\end{equation}}
\newcommand{\bc}{\begin{center}}
\newcommand{\ec}{\end{center}}
 \section{INTRODUCTION}

It is universally acknowledged that magnetization is crucial to the
accretion mechanism in circumstellar disks via the magneto-rotational 
instability 
(MRI) first
studied in the nonlinear regime by Hawley \& Balbus~(1991; see Balbus \& 
Hawley~1998 for a review).  Since
this accretion is the process by which most stars accumulate their masses from 
the
gravitational infall of collapsing, rotating, molecular cloud cores (see, e.g., 
Shu, Adams, \& Lizano
1987; Evans 1999), and since planets are believed to form from the resulting 
circumstellar 
disks (e.g,
Lissauer 1993; Lin \& Papaloizou 1995, 1996; Goldreich, Lithwick, \& Sari 2004), 
a better understanding of the 
mechanism of disk accretion (e.g., Lynden-Bell \& Pringle 1974, Pringle 1981)
is desirable for further progress in the fields of star and planet formation.
Moreover, bipolar outflows and jets are ubiquitous in young stellar objects 
(YSOs; see Bachiller 1995, Reipurth \& Bally
2001),
while the best contemporary theories for the underlying collimated winds 
intimately
involve the combination of rapidly rotating disks and strong magnetic fields, 
either threading the
disk itself or belonging to the central host star (K\"onigl \& Pudritz 2000, Shu 
et al. 2000).
A major uncertainty in the former models is the strength and geometry of the 
disk
magnetic fields.

In this paper, we consider the global problem of the mass,
angular-momentum, and magnetic-flux redistribution in disks around
young stars which are threaded by interstellar fields dragged in by the
process of gravitational collapse and infall. In a future extension of this 
work, we wish to include
the interaction of such magnetized accretion disks at their inner edges
with the stellar magnetosphere generated by dynamos operating in the
central objects.  Such interactions include the loss of angular momentum carried 
in any outflowing wind that develops at the surface
of the parts of the disk that rotate sufficiently close to Keplerian rotation, a 
process that is examined in this paper only
in terms of whether a wind's presence is implied by the prevailing circumstances. While separate pieces of 
this problem have been
attacked by other groups (e.g., Goodson, Bohm, \& Winglee 1999; Krasnopolsky \& 
K\"onigl 2002; K\"uker, Henning, \& R\"udiger 2004; Long, Romanova, \& Lovelace 2005), our study includes for the 
first time a perspective that combines analytical calculations with
the likely levels of magnetic field brought into the disk of a YSO by 
gravitational collapse
and infall (Galli et al. 2006; Shu et al. 2006).  

Our paper builds on the prescient study
of Lubow, Papaloizou, \& Pringle (1994, hereafter LPP) concerning the 
possibility of disk winds in accretion disks.
What distinguishes our work from theirs is our concern with the much stronger 
magnetic fields resulting from the process of star formation than assumed by LPP 
(see \S 3). Although LPP use a kinematic approximation that assumes explicitly 
an inward drift speed and implicitly a Keplerian rotation curve, neither 
assumption turns out to affect the generality of their solutions of the linear 
induction equation.  Nevertheless, when fields are dynamically strong, a 
solution of the full
magnetohydrodynamic (MHD) problem is required.  Thus, an added component of our 
study is a physical formula for the viscosity associated
with the MRI (see \S\S 2.1 and 4.1). In other words, while LPP treat magnetic 
fields as a passive contaminant to an imposed
accretion flow, we regard them as underlying the MRI dynamics that drives disk 
accretion.
Our own theoretical work on
the interaction of an electrically conducting accretion disk with a YSO
magnetosphere that produces an X-wind and a funnel flow
is incomplete because we previously ignored the magnetization of the accretion 
disk (see the review of Shu et al. 1999 and the criticisms of Ferreira, 
Dougados, \& Cabrit 2006).  

\subsection{Governing Equations}

We wish to calculate the effects of a systematically oriented, 
poloidal, mean magnetic field 
gathered from the interstellar medium that threads vertically through a 
circumstellar disk that surrounds a newly born star. 
This field is pinched radially inward by
viscous accretion through the thin disk driven by the MRI (Fig.~1).  To 
transform the usual
2.5 D equations of non-ideal MHD (with steep vertical stratification in $z$ 
combined with full radial dependences in $\varpi$ and axisymmetric motions in 
the tangential direction $\varphi$) to 1.5 D (integration over $z$) requires 
that we explicitly treat the vacuum fields above and below the disk all the way 
out to infinity, which is a 
crucial missing ingredient in all numerical simulations of the MRI to date when 
global fields are present.
Fortunately, this transformation can be implemented using 
the Green's function technique used by van Ballegooijen (1989) and LPP
(see also Shu \& Li 1997 and Shu, Li, \& Allen 2004, who were unaware of the
earlier related work on accretion disks until the
preparation of the present paper). 
In 1.5 D,
the formulation in terms of integro-differential
equations is then standard. 

Terquem (2003; see also Fromang, Terquem,
\& Nelson 2005), made the interesting suggestion that toroidal magnetic fields 
in YSO disks might be strong enough to stop the so-called Type I migration of 
planets and planetary embryos with Earthlike masses.   The origin of such mean 
toroidal fields is unclear since they require unclosed $z$-currents and are 
subject to buoyant vertical loss through the Parker (1966) instability, but 
similar effects could arise for accretion-pinned poloidal distributions.  The 
toroidal fields that arise in this paper from the stretching of radial fields by 
differential rotation vanish in the mean when we integrate over $z$.  We shall 
treat their fluctuating effects on the turbulent transport of angular momentum 
and matter across field lines as diffusive terms in the non-ideal equations of 
MHD with ``anomalous'' values for the coefficients of kinematic viscosity $\nu$ 
and electrical resistivity $\eta$ (see \S\S 2.1 and 4.1).

In such a mean-field MHD treatment, the evolution of gas and magnetic field
occurs in a thin axisymmetric, viscously accreting, disk 
surrounding a young star that we represent as
a stationary and gravitating point of mass $M_*$ at the origin of a cylindrical 
coordinate system $(\varpi, \varphi, z)$.
We denote the disk's surface density by $\Sigma$, the radial velocity of 
accretion in the plane by $u$, the angular velocity
of rotation about the $z$ axis by $\Omega$, the component of the magnetic field 
threading vertically through the disk by $B_z$,
and the radial component of the magnetic field just above the disk that responds 
to the radial accretion flow by $B_\varpi^+$.
This self-gravitating, magnetized system satisfies the time-dependent equation 
of continuity,
\be
{\partial \Sigma \over \partial t}+{1\over \varpi}{\partial \over \partial 
\varpi}\left( \varpi \Sigma u\right) = 0,
\label{eqcont}
\ee
the equation of radial-force balance,
\be
-\varpi \Omega^2 =  {B_zB_\varpi^+\over 2\pi \Sigma} -{1\over \varpi^2}\left[ 
GM_* +2\pi \int_0^\infty K_0\left( {r\over \varpi}\right)
G\Sigma (r,t) \, r dr\right],
\label{centribal}
\ee
the torque equation, including phenomenologically the effect of turbulent 
viscous stresses ($\propto \nu$),
\be
 \Sigma \left( {\partial \over \partial t} +u{\partial \over \partial 
\varpi}\right)\left( \varpi^2\Omega\right) =
 {1\over \varpi}{\partial \over \partial \varpi}\left( \varpi^2 \Sigma \nu 
\varpi {\partial \Omega\over \partial \varpi}\right),
 \label{torque}
 \ee
 and the induction equation for the vertical component of the magnetic field, 
including the effect of finite resistivity $\eta$,
 \be
 {\partial B_z \over \partial t}+{1\over \varpi} {\partial \over \partial 
\varpi}\left( \varpi B_z u\right) = -
 {1\over \varpi}{\partial \over \partial \varpi}\left( {\varpi \eta \over 
z_0}B_\varpi^+\right),
 \label{indeq}
 \ee
 where, according to Shu \& Li (1997),
 \be
 B_\varpi^+ = \int_0^\infty K_0\left( {r\over \varpi}\right) B_z(r,t)\, {r 
dr\over \varpi^2}.
 \label{radfield}
 \ee

In the vertical averaging over the thickness of the disk to arrive at 
equation (\ref{indeq}), 
we have effectively replaced $\eta_{\rm local} \partial B_\varpi/\partial z$ by 
its mean value above the mid-plane of the disk, $\eta B_\varpi^+/z_0$, an 
operation which defines
what we mean by average $\eta$.    We shall later consider what we mean by the 
effective half-height of the disk, $z_0$
(see Appendix C), but for the time being we are content with the intuitive 
concept.  Although it 
might be mathematically more elegant to absorb the combination $\eta/z_0$ into 
a single variable denoted, say, by a symbol $\cal R$, we 
retain the more cumbersome notation to keep better contact with the conventional 
microphysics of electrical resistivity.  In any case, we assume that $z_0$ is 
much smaller than the local disk radius $\varpi$.  

In equation (\ref{centribal}), the first term on the right-hand side represents 
the mass per unit area, $\Sigma$, divided into the radial 
component of Lorentz force per unit area due to magnetic tension, $J_\varphi 
B_z/c$, where $J_\varphi$ is the current density integrated over the thickness 
of the disk and equals $c B_\varpi^+/2\pi$ by Ampere's law.  The second and 
third terms on the right-hand side represent the contributions to the force per unit mass
associated with the stellar gravity of point mass 
$M_*$ and the self-gravity of the gas of surface density $\Sigma$ in the disk.  
To lowest order in the aspect ratio, $z_0/\varpi \ll 1$, we have neglected the 
pressure forces of the matter and the magnetic field (see Shu \& Li 1997).  The 
important astrophysical point is that the centripetal acceleration on the 
left-hand side of equation (\ref{centribal})  is not given {\it a priori} but 
arises in response to the forces of (1) magnetic tension,
(2) stellar gravity, (3) self-gravity of the disk gas, and (4) gas and magnetic 
pressure forces.  We have ignored (4) explicitly, and we shall presently ignore 
(3) also.  The rationale is that (3) and (4) are generally small in comparison 
with the stellar gravity term.   Their inclusion would only yield small 
corrections at the expense of rendering the resulting problem intractable except 
by numerical attack.  In contrast, the magnetic tension force 
is always present at a level dictated by the amount (and distribution) of 
magnetic flux threading the disk.

The kernel in equation(\ref{radfield}) is given by
\be
K_0(\xi)\equiv \frac{1}{2\pi}\oint\frac{1-\xi\cos\varphi}{(1+\xi^2-2\xi\cos\varphi)^{3/2}}\, 
d\varphi.
\ee
The function $K_0(\xi)$ is plotted in Figure 1 of Shu, Li, \&
Allen~(2004) with mathematical properties described in their Appendix~A. 
It has the asymptotic behaviors:
\be
K_0(0)=1 \mbox{~~and~~} K_0(\xi)\rightarrow -\frac{1}{2\xi^3} \mbox{~~as~~} 
\xi \rightarrow +\infty.
\ee

On a microscopic level, field diffusion in lightly ionized gases 
involves, in 
principle, three non-ideal effects (see Wardle \& Ng 1999): Ohmic resistivity 
(because ions are knocked off field lines by collisions), the Hall effect 
(because electrons move differently than ions under electromagnetic fields and 
collisions), and ambipolar diffusion (because neutrals do not feel 
electromagnetic forces directly, but are subject to them indirectly because of 
collisions with the ions).  For an axisymmetric problem, the Hall term vanishes, 
and the remaining two effects can be accorded the same treatment by defining an 
effective resistivity $\eta$ given by the sum of the Ohmic resistivity 
$\eta_{\rm Ohm}$ and the contribution from ambipolar diffusion $\tau B_z^2/4\pi 
\rho$,
where $\rho$ is a representative local volume density and $\tau$ is the mean 
collision time for momentum exchange between a neutral particle and a sea of 
charged particles or particulates (if charged dust grains are important):
\be
\eta = \eta_{\rm Ohm} + {\tau B_z^2 \over 4\pi \rho}.
\label {effresist}
\ee
For mean-field magnetohydrodynamics, we assume 
that the same relation applies
phenomenologically, and henceforth, when we speak of the ``resistivity,'' we 
mean the effective value (\ref{effresist}), appropriately generalized to include 
turbulent fluctuations.  In Appendix A, we comment on some alternative 
formulations
of the turbulent diffusive processes that result in the same conclusions when 
the system is in steady state.

\subsection{Steady State}

In steady state with a spatially constant mass-accretion rate $\dot M_{\rm d}$, the 
equations of continuity, torque, and advection-diffusion of magnetic field 
simplify to
 \be
 \varpi \Sigma u = {\rm const} \equiv -{\dot M_{\rm d} \over 2\pi},
 \label{Mdot}
 \ee
\be
\dot M_{\rm d} \varpi^2 \Omega = -2\pi \varpi^3 \Sigma \nu {d\Omega\over d\varpi}
\label{angmomtransf}
\ee
\be
B_z u  = -B_z{\dot M_{\rm d}\over 2\pi \varpi\Sigma} = -{\eta B_\varpi^+ \over z_0},
\label{fieldslippage}
\ee
whereas the centrifugal balance and the radial magnetic field $B_\varpi^+$ just 
above the disk plane are still given by equations (\ref{centribal}) and 
(\ref{radfield}) with no modifications except that there 
is no time dependence in the arguments of $\Sigma$ and $B_z$.
Elimination of $\dot M_{\rm d}$ from equations (\ref{angmomtransf}) and 
(\ref{fieldslippage}) then yields the self-consistency requirement,
\be
\eta {B_\varpi^+\over z_0 B_z} = -{\nu \over \Omega} {d\Omega\over d\varpi}.
\label{selfconsist}
\ee

In choosing the integration constants as above, we are implicitly 
allowing the origin (the star) to be a sink for mass but not for magnetic flux 
(or for angular momentum).  To make this clearer, multiply equation 
(\ref{indeq}) by $2\pi\varpi \, d\varpi$ and integrate in radius from the origin 
to a position just outside the star $R_*^+$ (which we will ultimately let 
$\rightarrow 0^+$).  The result yields
\be
{d\Phi_*\over dt} = -2\pi R_* \left( B_z u +{\eta B_\varpi^+ \over 
z_0}\right)_{\varpi = R_\ast^+} ,
\label{dPhi_*dt}
\ee
where $\Phi_\ast$ is the magnetic flux accreted by the star.
The assumption that equation (\ref{fieldslippage}) holds outside the star, in 
which the radial advection of magnetic field is everywhere balanced by the 
radial diffusion associated with diffusive effects, then implies that
\be
{d\Phi_*\over dt} = 0;
\label{nostellarflux}
\ee
i.e., all the magnetic flux that is brought in by the gravitational collapse 
involved in star formation is contained in the disk, if the star plus disk forms 
a closed system.  We emphasize that equation (\ref{nostellarflux}) is 
an {\it assumption}, not a deduction from first principles.  It is astronomical 
observations, not theory, that tell us that little interstellar flux is brought 
inside stars. Indeed, Mestel \& Spitzer (1956)
already recognized more than fifty years ago that stellar fields would measure
in the megagauss range if even a small fraction of the original interstellar
flux were to appear on a young star's surface (see also Shu et al.~2004, 2006).

In this context, it is important to realize that magnetic reconnection will only destroy field lines that close within the disk. Indeed, the annihilation of field loops is essential for fluid transport through the disk, as discussed in \S 4.1 below. In contrast, the open, mean, magnetic fields pictured in Figure 1 have roots that extend to the interstellar medium. These field lines are responsible for the flux (integral of $B_z$ over the area of the disk), and this flux cannot be lowered without modifying the interstellar currents (assumed to vanish at all finite distances in the derivation of eq.~[\ref{radfield}]), for example, by arbitrarily adding spatially large loops of current that yield a flux of opposite sign to the original interstellar value.\footnote{One might still worry about the fate of open field lines with a fixed flux that connects a rapidly rotating disk with a slowly rotating interstellar cloud. Wouldn't such field lines get twisted up in time and torque down the disk? The answer is yes, but for disturbances traveling a poloidal distance $\Delta s$ along a field line out of the plane of the disk, the change in azimuthal angle $\Delta \varphi$ experienced by a loaded field line is approximately given by the disk rotation rate times the Alfv\'en crossing time across the region: $\Delta \varphi \sim \Omega(\Delta s /v_A)$. During the infall stage, the mass loading is so considerable that the ability even to form a disk is in jeopardy (see, e.g., Galli et al.~2006). After infall has stopped, the Alfv\'en speed $v_A$ is very large near but above the disk, and it decreases as the field line begins to penetrate the interstellar cloud. In this stage, all the twist is at the cloud end; virtually none is at the disk end.  In other words, equation (\ref{fieldratio}) below will hold to good approximation.  There is little continued torquing of the disk because the rate of angular momentum transport per unit area by torsional Alfv\'en waves is limited by the angular-momentum density $\rho \varpi^2\Omega$ times the Alfv\'en speed $v_A = B/\sqrt{4\pi \rho}$, with the small $\rho$ above the disk of the former more than canceling the $\sqrt{\rho}$ in the denominator of the latter.} 
 ÊÊ

In what follows, then, we choose to write equation (\ref{selfconsist}) as
\be
B_\varpi^+ = \alpha^2 B_z, 
\label{fieldscale}
\ee
where we define the dimensionless auxiliary variable $\alpha$ by
\be
\alpha^2 \equiv -{z_0\nu \over \varpi \eta}\left( {\varpi \over \Omega} 
{d\Omega \over d\varpi}\right).
\label{auxvar}
\ee
For realistic configurations, we expect $B_\varpi^+\sim B_z$, i.e., we 
anticipate that $\alpha$ is an order unity quantity. Since $\varpi 
\Omega^{-1}d\Omega/d\varpi$ is also of order unity, the resistivity $\eta$ must 
be smaller than the (turbulent) viscosity $\nu$ by a factor given roughly by the 
local aspect ratio of the disk, i.e.,
\be
{\eta \over \nu} \sim {z_0\over \varpi} \ll 1.
\label{ordermag}
\ee
Equation (\ref{ordermag}) is in agreement with the assertion in equation (39) of 
LPP that the dimensionless ratio of interest
is not $\eta/\nu$ but $(\eta/\nu)(\varpi/z_0)$, which must be of order unity for 
magnetic fields to
be bent by an order unity amount from the vertical direction.  The contrary 
assertion by R\"udiger \& Shalybkov (2002) arises
because they arbitrarily assume a uniformly-rotating halo of conducting matter 
in which the differentially
rotating disk is embedded.  Some aspects of the problem that R\"udiger \& 
Shalybkov consider could apply to the interaction
of the inner edge of an accretion disk with a uniformly rotating stellar 
magnetosphere (e.g., K\"uker et al. 2004).  The latter
is the context of X-wind theory.
In the next subsection, we give an explicit justification of the 
order-of-magnitude estimate in equation (\ref{ordermag}) when the disk mass is 
small enough to allow us to ignore its self-gravity.  

\subsection{Exact solution when self-gravity of the disk is neglected}

If we may ignore the last term (disk self-gravity) in equation (\ref{centribal}) 
in comparison to the other terms on the right-hand side (magnetic tension and 
stellar gravity), then the condition of centrifugal force balance becomes
\be
\varpi \Omega^2 = - {B_zB_\varpi^+\over 2\pi \Sigma} +{GM_*\over \varpi^2},
\label{centrifugal}
\ee
where $B_\varpi^+$ is given by equation (\ref{radfield}). 
Consider now the important case when $\Omega$ is appropriate for a thin disk in 
quasi-Keplerian rotation,
\be
\Omega = f \left({GM_*\over \varpi^3}\right)^{1/2}, 
\label{quasikep}
\ee
where $f$ is a constant less than 1 because of partial support of the disk 
against the stellar gravity by the magnetic tension of the poloidal magnetic 
fields that thread through it, brought into the system by the process of star 
formation. 
In order for these conditions to be mutually compatible, equation 
(\ref{centrifugal}) requires
\be
B_z = \alpha^{-1}\left[ 2\pi (1-f^2)GM_* \varpi^{-2}\Sigma \right]^{1/2}.
\label{Bz}
\ee
The substitution of equation (\ref{Bz}) into equations (\ref{fieldscale}) and 
(\ref{radfield}) now results in a linear integral equation for $\Sigma^{1/2}$ 
when $\alpha$ is known,
\be
\alpha (\varpi)\Sigma^{1/2}(\varpi) = \int_0^\infty K_0\left( {r\over 
\varpi}\right) \alpha^{-1}(r)\Sigma^{1/2}(r)  \, {dr \over \varpi}.
\label{Sigmainteq}
\ee

Because $\Sigma^{1/2}$ enters linearly into equation (\ref{Sigmainteq}) and 
$\alpha$ enters nonlinearly, the above relationship yields a constraint on the 
allowable solutions for both $\Sigma$ and $\alpha$. Consistent with equation 
(\ref{quasikep}), suppose, for example, that 
$\Sigma$ is given by a power law:
\be
\Sigma(\varpi) = C \varpi^{-2\ell},
\label{powerlawS}
 \ee
where $C$ and $\ell$ are constants.  Then, equation (\ref{Sigmainteq}) requires 
$\alpha^2$ to be a positive dimensionless constant:
\be
\alpha^2 = I_\ell \equiv \int_0^\infty K_0(\xi) \xi^{-\ell} \, d\xi.
\label{alpha2}
\ee
The integral $I_\ell$ has a finite value for $\ell$ between $-2$ and 1 with $I_0 
= 1$.  Table 1 gives a tabulation of numerical values for the astronomically 
interesting range of $\ell$ from 0 to 1 where $\Sigma$ is a declining function 
of radius.  The last row shows the inclination angle $i$ that the surface field 
makes
with respect to the vertical direction as computed from $\tan i = 
B_\varpi^+/B_z$ and
equation (\ref{fieldratio}) below.  

\begin{deluxetable}{lllllllllllllll}
\rotate
\tablecolumns{15}
\tablewidth{0pc}
\tablecaption{Values of $I_\ell$ and inclination angle $i$}
\tablehead{}
\startdata
$\ell$     & 0 & 0.1 & 0.2 & 1/4 & 0.3 & 5/16 & 3/8 & 0.4 & 1/2 & 0.6 & 0.7 & 0.8 & 0.9 & 1  \\
$I_\ell$   & 1 & 1.149 & 1.326 & 1.428 & 1.542 & 1.573 & 1.742 & 1.818 & 2.188 & 2.726 & 
3.598 & 5.304 & 10.34 & $\infty$ \\
$i$        & $45^\circ$ & $49.0^\circ$ & $53.0^\circ$ & $55.0^\circ$ & 
$57.0^\circ$ & $57.5^\circ$ & $60.1^\circ$ & $61.2^\circ$ & $65.4^\circ$ & $69.9^\circ$ & 
$74.5^\circ$ & $79.3^\circ$ & $84.5^\circ$ & $90^\circ$ \\
\enddata 
\end{deluxetable}

Equations (\ref{alpha2}) and (\ref{quasikep}) now allow us to deduce from 
equation 
(\ref{auxvar}) the
required relationship between the resistivity and viscosity as
\be
{\eta \over \nu} = {3\over 2I_\ell}\left( {z_0\over \varpi}\right),
\label{Prandtl}
\ee
which supplies the missing numerical coefficient to our previous estimate 
(\ref{ordermag}).  The Prandtl combination $\eta/\nu$ is required in a 
steady-state accretion disk
to have the specific ratio (\ref{Prandtl}) if $u =-3\nu/2\varpi$ arising from 
the viscous transport of angular momentum (cf.~eqs.~[\ref{Mdot}] and 
[\ref{angmomtransf}])
is the same drift speed needed for the resistive diffusion of matter across 
stationary field lines, $u = -(\eta/z_0) B_\varpi^+/B_z$ (cf.~eq.~[\ref{fieldslippage}]).  
The two formulae for the drift velocity express succinctly  why $\eta$ only 
needs to be a small fraction of $\nu$: unlike viscosity, resistivity is not 
acting to mix quantities on a large scale of $\varpi$; instead, it is trying to 
annihilate the oppositely-directed mean radial-fields $B_\varpi$ on either side 
of the mid-plane distributed on a small scale $z_0$.  If we consider the 
diffusivity associated with the {\it radial} distribution of the current 
$J_\varphi$ (Appendix A), then that diffusivity $\eta_J$ {\it is} approximately 
equal to $\nu$.  
  
The corresponding relationship between the radial component of the magnetic 
field at the upper surface of the disk and the vertical component at the 
mid-plane is given by equation (\ref{fieldscale}) as
\be
B_\varpi^+ = I_\ell B_z.
\label{fieldratio}
\ee
The vertical field is itself given by equation (\ref{Bz}) when we know the 
surface density from equation (\ref{powerlawS}), i.e.,
\be
B_z = I_\ell^{-1/2}\left[ 2\pi (1-f^2)GM_*C \right]^{1/2}\varpi^{-(1+\ell)}.
\label{vertfield}
\ee
To obtain the identification of the coefficient $C$, we note that equation 
(\ref{angmomtransf}) implies
\be
\Sigma = {\dot M_{\rm d}\over 3 \pi \nu}.
\label{viscousSig}
\ee
Thus, the adoption of (\ref{powerlawS}) is an implicit assumption that the 
viscosity varies as a power-law of $\varpi$ in steady state given by
\be
\nu = \left({\dot M_{\rm d}\over 3\pi C}\right)  \varpi^{2\ell}.
\label{powerlawnu}
\ee 

To make further progress, we need to have a physical theory for the kinematic 
viscosity $\nu$ (see \S\S 2.1 and 4.1). This will allow the last unknown 
quantities $\ell$ and $C$ to be eliminated 
from our power-law solution for $\Omega$, $\Sigma$, and $B_z$.  We discuss 
first, however, in \S\S 1.4 and 1.5 two results that hold for more general 
diffusivities.

\subsection{Disk winds}

There are two criteria necessary for a cold wind to be magnetocentrifugally 
driven from
the surface of a rotating disk (cf. the review of K\"onigl \& Pudritz 2000).
Without detailed specifications of the physics of the viscosity or the 
resistivity, equation (\ref{fieldratio}) allows us to confirm the conclusion 
reached by LPP 
concerning the first criterion: 
\be
\frac{B_\varpi^+}{B_z} = I_\ell \ge\frac{1}{\sqrt{3}},
\label{windcrit}
\ee
so that the footpoint field would then bend with an inclination angle $i$ from 
the vertical by more than
$30^\circ$ (Chan \& Henriksen 1980, Blandford \& Payne 1982). Table 1
shows that the above criterion is comfortably satisfied for any disk where the
surface density declines with radius, $\ell > 0$. In particular, $\ell = 1/4$ in equation (\ref{vertfield}) corresponds to 
the famous case $B_z \propto \varpi^{-5/4}$ considered by Blandford \& Payne (1982) and yields $i = 55.0^\circ$.
LPP state their wind-launching criterion in the form
that $(\eta/\nu)(\varpi/z_0)$, which equals $3/(2I_\ell)$ according to equation 
(\ref{Prandtl}),
should be less than $1.52\sqrt{3}$.  This result is almost identical to the 
criterion (\ref{windcrit}).
LPP's calculation is for a finite
disk with nonzero inner and outer radii, embedded in a background field of 
uniform strength pointing in the vertical direction, whereas our calculation is 
formally for an infinite disk with a trapped interstellar flux.  The negligible 
astrophysical difference between 1.52
and 3/2 implies that none of these idealizations matter to the first criterion 
for driving a disk wind.

Unfortunately, the satisfaction of the magnetic criterion 
(\ref{windcrit})
by itself is not a sufficient condition for the appearance of a
significant disk wind.  Equation (\ref{A2}), derived from the 
consideration of vertical hydrostatic equilibrium, shows that if the 
fractional departure from Keplerian rotation $1-f^2$ is large in 
comparison with the aspect ratio, $A \equiv 
z_0/\varpi$ divided by $I_\ell$, then the square of the characteristic thermal speed in the 
disk interior,
$a^2$, is given by
\be
a^2 \approx {I_\ell\over 2}A(1-f^2){GM_*\over \varpi}.
\label{interior}
\ee
On the other hand, in order to drive a disk wind, the square of the 
thermal speed at the disk surface, $a_s^2$, must be greater
than a fraction (say, 1/4) times the virial imbalance between the 
gravitational potential and twice the specific kinetic energy in disk 
rotation:
\be
a_s^2 > {1\over 4} (1-f^2){GM_*\over \varpi}.
\label{diskwindcrit}
\ee
The 1/4 on the right-hand side has the following approximate 
justification.
Parker's solution for a thermally-driven spherical wind gives one 
factor of 1/2 (see Parker 1963); this 1/2 becomes 1/4 because a 
particle in Keplerian rotation already has 1/2 of the energy needed to 
escape.  No other factors are then included because magnetocentrifugal 
effects do not help MHD winds in making the sonic transition (see the 
discussion of Shu et al. 1994).

Except for the effects of heating by external irradiation, the square of the thermal speed $a_s^2$ of the gas at the disk 
surface is likely to be small in comparison with its value $a^2$ in the disk 
interior when the disk is vigorously accreting, the condition needed to 
allow equation (\ref{A2}) to hold.  The inequality (\ref{diskwindcrit}) 
is then inconsistent with equation (\ref{interior}), implying that 
strong disk winds cannot be driven when $1-f^2$ much exceeds the small 
number $A$ (the disk aspect ratio $z_0/\varpi$).  Otherwise, the 
surface that corresponds to smooth slow-MHD crossing would lie at so
many scale heights above some nominal disk surface that the associated
mass-loss rate would become negligibly small.  LPP avoided this problem
by their implicit assumption that the fields threading the disk were 
weak and therefore
had no effect on the disk's assumed Keplerian rotation.
Wardle \& K\"onigl (1993) examined the same issue in a {\it local} 
treatment of the launch region for
disk winds assuming ambipolar diffusion to be the physical mechanism 
that loads field lines.
They reported that the effect is present, but compensating factors 
exist that allow wind mass-loss rates
to be a small fraction of the disk accretion rate, with a very 
sensitive dependence on the ratio of the orbit time to the ion-neutral 
collisional time $\tau$ (see Fig.~12 of their paper).
The {\it global} treatment given in this paper, which includes an 
assessment of the
likely levels of magnetic field strength to result from the process of 
star formation,
indicates that the problem is more severe,
perhaps even insurmountable for FU Orionis and T Tauri stars, although 
the situation may
yet be rescued for the outer disk regions of embedded low- and 
high-mass protostars where $A$ is not so small (see \S 3).

Font et al. (2004) propose that photo-evaporation is a more likely source 
of the slow, warm disk-wind observed by Kwan et al. (2006) in T Tauri stars.  A photo-evaporative
wind could reach a higher terminal velocity, but not a
larger mass-loss rate (limited by the X-ray, EUV, or FUV photon-flux reaching the surface of the
disk at radii of a few AU or greater), because of the boost to the gas by magnetocentrifugal
fling after the sonic transition is made.  The combined effect could lead to a significant loss of angular momentum from the system in the late stages of YSO evolution that is not taken into account in our treatment.  In other words,
{\it outer} disk winds may realistically arise even if $1-f^2$ is not small. 

\subsection{Viscous and resistive dissipation rates}

In steady state, the net emergent radiation from the upper and lower surfaces of 
a thin disk, after accounting for the irradiation of the central star, has to 
carry away the sum of the energies generated by viscous and resistive 
dissipation, whose rates per unit area are:
\be
\Psi = \nu\Sigma
\left(\varpi {d\Omega\over d\varpi}\right)^2 ={3\over 2}f^2\left( {GM_*\dot 
M_*\over 2\pi \varpi^3}\right);
\label{viscdiss}
\ee
\be
Y\equiv J_\varphi E_\varphi = \left({cB_\varpi^+\over 2\pi}\right)\left(- 
{u\over c}B_z\right) =
(1-f^2){GM_*\dot M_{\rm d}\over 2\pi \varpi^3},
\label{resistdiss}
\ee
In the above, $J_\varphi = cB_\varpi^+/2\pi$ and $E_\varphi = -uB_z/c$ are, 
respectively, the mean current density (integrated over $z$) and electric field 
in the $\varphi$ direction in the rest frame of the
plasma tied to mean $B_z$ relative to which the bulk of the matter in the disk 
is drifting at radial velocity $u$ =
$-\dot M_{\rm d}/2\pi \varpi \Sigma$ = $-3\nu/2\varpi$ = $-(\eta/z_0) B_\varpi^+/B_z$.   
To perform the last step
in equation (\ref{resistdiss}), we have used equations (\ref{centrifugal}) and 
(\ref{quasikep}) to eliminate $B_\varpi^+B_z/2\pi$ and $\Omega$.

An alternative expression, $Y = \eta {(B_\varpi^+)^2/2\pi z_0}$, makes more 
apparent that $Y$ represents
the resistive dissipation, which feeds on the magnetic tension.  In contrast, 
the viscous dissipation $\Psi$ feeds on the disk shear.
In the former case,  a Lorentz force drives electric currents that 
generate heat by friction between the various charged and non-charged species; 
in the latter,
heat is generated by fluid elements "rubbing" tangentially against
each another.
We speculate that a fraction of the energy released by
resistive dissipation in the disk may go into accelerating suprathermal 
particles that give proto-planetary disks
higher ionization rates than conventionally estimated (cf.~Goldreich \& 
Lynden-Bell 1969 vs.~Sano et al. 2000).

The coefficient $3f^2/2$ in equation (\ref{viscdiss}) differs from the standard 
result by the factor $f^2$ 
because the rotation law is only quasi-Keplerian, and implies a local rate of 
energy release 3 times as great as one might have expected from the loss of 
orbital energy because of accretion (see, e.g., Lynden-Bell \& Pringle 1974).  
The difference is made up by a viscous torque that transfers energy from the 
inner disk to the outer disk, a debt that has to be repaid if we were to examine 
the details of the interaction of the disk near its inner edge with a star of 
finite size and, perhaps, magnetization.  For example, if one applies a zero 
torque condition at an inner boundary $R_{\rm x}$
corresponding to a stellar magnetopause that corotates at the angular rate 
$\Omega(R_{\rm x}) = f(GM_*/R_{\rm x}^3)^{1/2}$,
then standard arguments (Lynden-Bell \& Pringle 1974) show that the total 
viscous energy dissipated by bringing matter through our accretion disk from 
infinity to $R_{\rm x}$ equals 
\be
{1\over 2}f^2{GM_*\dot M_{\rm d}\over R_{\rm x}}.
\label{viscoustotal}
\ee

In contrast, if we multiply equation (\ref{resistdiss}) by $2\pi\, \varpi 
d\varpi$ and integrate from $R_{\rm x}$ to $\infty$,
we derive that the rate of resistive dissipation of energy in the disk equals
\be
(1-f^2){GM_*\dot M_{\rm d} \over R_{\rm x}} .
\label{resistivetotal}
\ee
The sum of the viscous and resistive dissipation rates, equations 
(\ref{viscoustotal}) and (\ref{resistivetotal}), equals
\be
\left(1 -{1\over 2}f^2\right){GM_*\dot M_{\rm d} \over R_{\rm x}},
\ee
which is not the total rate of gravitational potential energy release, $GM_*\dot 
M_{\rm d}/R_{\rm x}$, because an amount $R_{\rm x}^2\Omega^2(R_{\rm x})/2 = 
f^2GM_*/2R_{\rm x}$ is still retained by each gram of disk matter at the 
disk/stellar-magnetopause boundary as specific orbital energy.  In actual practice, as will be shown in
a future publication, $f$ is raised to unity as the latter boundary is crossed
by the swing of $B_\varpi^+B_z$ from positive to negative values and inner-edge
effects (assuming the magnetopause
is not squashed by the accretion flow to the stellar surface in quasi-steady state),
so $GM_*/2R_X$ of specific energy
in the accreting matter is available for budgeting in a funnel flow or X-wind (Shu 1995).
Notice that in this description, electromagnetic fields, although 
responsible for the microphysics of viscous and resistive dissipation, act on 
the macroscale merely as catalysts for converting gravitational energy into 
other forms.  These other forms, in steady state, involve no change of the 
magnetic energy because the magnetic fields have been assumed to remain constant 
in time.

Let us compare the expressions (\ref{viscdiss}) and (\ref{resistdiss}) which 
hold at radii $\varpi \gg R_{\rm x}$.  Then we easily calculate that heat 
generation by viscous dissipation dominates over resistive dissipation when $f > 
\sqrt{2/5} = 0.6325$.  The 
resistive contribution is typically not negligible; for example, it is 37.5\% of 
the viscous contribution when $f=0.8$.
In spirit, if not in detail, our ideas then follow those of Lynden-Bell (1969), 
and we can
imagine ``resistive accretion disks'' as well as their ``viscous'' counterparts 
(cf. the FU Orionis 
model of \S 3).

\section{VISCOSITY ASSOCIATED WITH MRI}
 
Returning to our quest for the viscosity $\nu$ to be used in our
power-law solution in \S 1.3, we would like to benefit from the 
many numerical simulations that have been performed of the MRI since 
the pioneering studies of Balbus and Hawley. However, few experiments
have been done that are of direct relevance to the problem in
star formation that we address in this paper.  Most simulations miss one or the 
other of the 
crucial ingredients of being both {\it global} and having {\it nonzero net 
flux}.  When field lines extend to infinity (necessary
to have nonzero net flux), rather than close within the system, consideration of 
the behavior of the field within a few vertical scale heights of a spatially 
thin disk suffices only if one has included knowledge of what those fields do at 
large distances.  The application of boundary conditions at smaller distances 
will generally exert extraneous stresses.  
Because of these difficulties, many simulations are both local and have zero net 
flux,
in that a small portion of a shearing sheet or layer is threaded by mean 
vertical magnetic fields $B_z$
inputted initially to vary sinusoidally in the radial direction.  
Radial mixing and reconnection can destroy most of the initial vertical 
field in such simulations, so that the turbulent state reached asymptotically
in time is largely independent of the assumptions of the ``initial''
state.  The finely resolved study by Silver \& Balbus (2006) does include 
the effect of a systematically directed field $B_z$ of a single sign, but their 
simulation is not global
(and therefore does not develop a $B_\varpi^+$ comparable to $B_z$), and it is 
performed
for a gas pressure 800 times larger than the magnetic pressure, i.e., with 
implied magnetic
fields that are too weak to be useful for our studies here. 

Stone et al. (1996) and Miller \& Stone (2000) performed, to our 
knowledge, the only well-known simulations of a thin disk threaded by a 
systematic, large-scale, nonzero, vertical field $B_z$.  The computations were 
semi-global in spanning a
larger than usual, but still limited, range of $z$.  (The work by 
Fleming, Stone, \& Hawley 2000, assumes periodicity in the $z$ direction, 
which does not faithfully represent the dynamics of a thin disk.)  The
thin-disk cases with non-zero net flux behave completely differently from 
the other more frequently studied configurations, whose initial states 
have only toroidal fields, or, at least, zero average $B_z$.  The 
systems in the simulations of Stone et al. (1996) and Miller \& Stone 
(2000) quickly become magnetically dominated, unlike the usually considered
circumstance where the gas pressure is much greater than the magnetic 
pressure.  The rapid evolution then prevented the authors from examining the 
astrophysical 
consequences of the configuration most likely to be relevant to investigations 
in star formation.  

\subsection{Turbulent Viscosity}

In the absence of relevant numerical simulations, we give the following order of 
magnitude argument on the basis of mixing-length ideas (Prandtl 1925).
When a radial field $B_\varpi$ is present in a field of differential rotation, 
we expect that field to be sheared
and yield an azimuthal component $B_\varphi$.  The tendency for electrically 
conducting fluids to flow along the
field direction suggests that fluctuations in the radial velocity $\delta u$ 
will be related
to the horizontal fields and the shear rate by
\be
B_\varphi\delta u \sim B_\varpi \varpi {d\Omega\over d\varpi}\delta \varpi,
\label{horizfieldfreeze}
\ee
where $\delta \varpi$ is the radial mixing length and has the same sign as 
$\delta u$.  Notice that the induced
$B_\varphi$ has systematically the opposite sign as $B_\varpi$ if 
$d\Omega/d\varpi$ is negative.  The systematics of
$B_\varphi$ relative to $B_\varpi$ lead to the desired ``viscous'' torque.  

Differentially rotating fluid parcels displaced from their equilibrium positions
that preserve their specific angular momentum gyrate in epicycles about a 
guiding center
characterized by an epicyclic frequency $\kappa$ that numerically equals
$\Omega$ in a quasi-Keplerian disk (Binney \& Tremaine 1987).
Although other forces are also at play in a magnetized accretion disk, we
assume that mixing-length scales of greatest interest for the transport of 
angular momentum
have a correlation time
between $\delta \varpi$ and $\delta u$ that is similarly given by $\sim 
\Omega^{-1}$, i.e.,
\be
\delta u \sim \Omega \delta \varpi .
\ee
Equation (\ref{horizfieldfreeze}) can now be written
\be
B_\varphi \sim B_\varpi {\varpi\over \Omega}{d\Omega \over d\varpi}.
\label{Bphi}
\ee
The component of Maxwell stress responsible for exerting torque, 
$B_\varpi B_\varphi/4\pi$,
integrated over the thickness of the disk, can then be approximated as
\be
{\cal F}{(B_\varpi^+)^2\over 2\pi} {\varpi\over \Omega} {d\Omega\over 
d\varpi}z_0.
\label{Maxwell}
\ee
where ${\cal F}$ is a form factor that comes from the vertical integration, and 
that  
also corrects for all the order-of-magnitude approximations used to arrive at 
this point.
If the term (\ref{Maxwell}) is the dominant contribution to the ``viscous'' 
stress modeled in equation (\ref{torque}),
then the associated ``kinematic viscosity'' equals
\be
\nu = {\cal F}{(B_\varpi^+)^2 z_0\over 2\pi\Sigma\Omega}.
\label{kinematicvisc}
\ee

A more picturesque ``derivation'' of equation (\ref{kinematicvisc}) which 
explains why the correlations do not
involve quadratic products of fluctuating quantities (they actually do) and why 
the mixing length $\delta \varpi$ seemingly dropped out of the calculation (it 
should not) is given in \S 4.1.  We proceed here to build confidence in the case 
by first demonstrating that the adoption of equation (\ref{kinematicvisc}) leads 
to reasonable astrophysical results.

To follow the flux redistribution, our problem is formulated using $B_z$ rather 
than $B_\varpi^+$, so in steady-state we set $B_\varpi^+ = I_\ell B_z$ 
(cf.~eq.~[\ref{fieldratio}]) and get
\be
\nu=D\frac{B_z^2 z_0}{2\pi \Sigma \Omega},
\label{def_nu}
\ee
where $D$ is a dimensionless coefficient given by $D = I_\ell^2 {\cal F}$.  
Although 
${\cal F}$ is the more fundamental quantity, as we shall see in a later study of 
the interaction of disks with stellar magnetospheres, we shall use $D$ in this 
paper as the relevant dimensionless parameter to obtain 
$\nu$ from observations and simulations
(see the discussion of Appendix B).  As long as ${\cal F}$ is not too small 
compared 
to unity, $D$ is an order unity quantity,
provided the entire disk layer undergoes vigorous mixing from the MRI.
 
With $\nu$ given by equation (\ref{def_nu}), equation (\ref{viscousSig}) yields 
the following expression for the vertical magnetic field:
\be
B_z= \left(\frac{2f}{3DA}\right)^{1/2}
\left(\frac{GM_*\dot M_{\rm d}^2}{\varpi^5}\right)^{1/4},
\label{bz_pd}
\ee
where we have used equation (\ref{quasikep}) to express the angular rotation 
rate $\Omega$ and
where $A\equiv z_0/\varpi$ is the aspect ratio of the local disk height to
disk radius.  From equations (\ref{Bz}) and (\ref{A2}), we are then able to 
recover the surface density as
\be
\Sigma=\frac{f}{1-f^2}
\left(\frac{I_\ell}{3\pi D A}\right)
\frac{\dot M_{\rm d}}{(GM_*\varpi)^{1/2}}.
\label{sig_pd}
\ee

Equation (\ref{def_nu}) holds with nonzero $D$ only as long as (1) good magnetic 
coupling exists, and 
(2) the criterion for the MRI instability is satisfied, that 
the magnetic pressure be smaller than the gas pressure.  If we define a fiducial 
square of the thermal velocity $a^2(\varpi)$ by the gas pressure at the 
midplane, $P(\varpi, 0)$, divided by the characteristic volume-density in the 
disk, $\Sigma(\varpi)/2z_0(\varpi)$, equations (\ref{bz_pd}) and (\ref{sig_pd}) 
imply that the ratio of the magnetic pressure to gas pressure at the disk's 
midplane is then given by
\be
{B_z^2 z_0\over 4\pi \Sigma a^2} = {(1-f^2)\over 2I_\ell}\left( {AGM_*\over 
a^2\varpi}\right),
\label{magpresstogaspress}
\ee
where we have expressed $z_0 = A\varpi$.  On the other hand, analysis of the 
vertical hydrostatic equilibrium of the disk using the method of Wardle \& K\"onigl (1993; see also Ogilvie 1997 
and Shu \& Li 1997) gives (see Appendix C):
\be
a^2 = {1\over 2}\left[ I_\ell (1-f^2)A+A^2\right]{GM_*\over \varpi}.
\label{implicit_a2}
\ee
In principle, a proper physical treatment would require us to obtain 
$a^2(\varpi)$ by computing the volumetric viscous and resistive heating as a 
function of $z$ and balance it against the heat transported by radiative 
transfer and thermal convection vertically out of the disk.  Once we have 
obtained $a^2(\varpi)$, we could then solve equation (\ref{implicit_a2}) as a 
quadratic equation for the disk aspect ratio $A$.  Such a physically involved 
treatment is beyond the scope of the present paper, and for the astronomical and 
pedagogical sake of obtaining numerical examples, we shall assume the luxury of 
specifying $A(\varpi)$ semi-empirically as a power-law (see \S 2.3).

The justification for approximating $A$ as a power-law follows.  If disks are 
spatially thin, $A$ is small compared to unity.  There are then two 
different regimes of physical interest.  The first case arises when
the departure $(1-f^2)$ from Keplerian rotation is small, and the first-term on 
the right-hand side of equation (\ref{implicit_a2}) is negligible in comparison 
with the second term, resulting in the approximation:
\be
A \approx a\sqrt{{2\varpi\over GM_*}} \qquad {\rm for} \qquad 1-f^2 \ll {A\over 
I_\ell}.
\label{A1}
\ee
In case (\ref{A1}), the contribution of the magnetic pressure 
is ignorable for the vertical hydrostatic equilibrium, and the draw of the 
stellar gravity toward the midplane keeps a cool accretion disk 
spatially thin.
The second case arises when the departure from Keplerian rotation $(1-f^2)$ is not small,
and the first-term on the right-hand side 
of equation (\ref{implicit_a2}) dominates over the second:
\be
A \approx {2\over I_\ell}\left[{a^2\varpi\over GM_*(1-f^2)}\right] \qquad {\rm 
for} \qquad 1-f^2 \gg {A\over I_\ell}.
\label{A2}
\ee
In case (\ref{A2}), the disk is kept spatially thin, not 
by stellar gravity, but by the inward press of the component of magnetic 
pressure $B_\varpi^2/8\pi$ which increases outward from the midplane $z=0$, 
where its value is 0, to the surface, where its value is $(B_\varpi^+)^2/8\pi = 
I_\ell^2 B_z^2/8\pi$.

From the community experience in disk thermal-physics, it is well-known that 
numerical solutions frequently show two types of power-law solutions for the 
vertically-averaged temperature $\propto a^2(\varpi)$, namely, $a^2 \propto 
\varpi^{-3/4}$ for passive, irradiated disks, and $a^2 \propto \varpi^{-1/2}$ 
for active, accretion-powered disks.  For $a^2 \propto \varpi^{-3/4}$, case 
(\ref{A1}) yields $A(\varpi) \propto \varpi^{1/8}$ and case (\ref{A2}) yields 
$A(\varpi) \propto \varpi^{1/4}$.   For $a^2 \propto \varpi^{-1/2}$, case 
(\ref{A1}) yields
$A(\varpi) \propto \varpi^{1/4}$ and case (\ref{A2}) yields $A(\varpi) \propto 
\varpi^{1/2}$.
Therefore, a power-law description for disk flaring, $A(\varpi) \propto 
\varpi^n$, with $n = 1/4$ typically or $n = 1/8$ or $1/2$ at the extremes, has a theoretical basis.

Apart from the issue of sufficient ionization, the requirement for MRI being 
present in the mid-plane is that the left-hand side of equation 
(\ref{magpresstogaspress}) should be less than unity. This requirement is {\it 
automatically} satisfied for all our disks because the substitution of equation (\ref{implicit_a2}) into 
the right-hand side of equation (\ref{magpresstogaspress}) gives a 
value $(1-f^2)/[(1-f^2)I_\ell^2+AI_\ell]$, which is always smaller than 1 for $I_\ell \ge 1$, i.e.,
for $\ell \ge 0$ (Table 1).  In particular, in the regime where equation (\ref{A2}) holds,
$\nu$ from equation (\ref{def_nu}) becomes
\be
\nu = {2D\over I_\ell^2} \left( {a^2\over \Omega}\right),
\label{Shakura-Sunyaev}
\ee
which has the same form as Shakura \& Sunyaev (1973) viscosity, $\nu \equiv 
\alpha_{\rm ss} a^2/\Omega$ with $\alpha_{\rm ss} = 2D/I_\ell^2$.  The strongly 
magnetized disks of this paper therefore both automatically satisfy the 
criterion that MRI exists in the midplane and have equivalent Shakura-Sunyaev 
alpha's of order unity if $D \sim 1$.  By compressing the midplane gas density
and pressure to higher values than gravity can achieve alone, such disks always 
operate at nearly the maximum efficiency for viscous transport, if $D \sim 1$, 
without shutting down the MRI.  Aficionados of MRI like to say that it is 
present in thin disks for arbitrarily low levels of magnetic field; now they can 
add that it is present for arbitrarily high values too, provided $I_\ell = 
B_\varpi^+/B_z \ge 1$.

For later reference in discussions of disk fragmentation, we record that the 
local dimensionless mass-to-flux ratio in the disk is given by (see Nakano \& 
Nakamura 1978, Basu \& Mouschovias 1994,  Shu \& Li 1997, Krasnopolsky \& Gammie 
2005):
\be
\lambda \equiv {2\pi G^{1/2}\Sigma \over B_z} =
{I_\ell\over 1-f^2}\left({2f\over 3 DA}\right)^{1/2}\left( {\dot M_{\rm d}^2 
\varpi^3\over GM_*^3}\right)^{1/4}.
\label{masstoflux}
\ee
The supercritical condition $\lambda > 1$ is necessary, but not sufficient, for 
local disk fragmentation. We must also examine, at least, the Toomre (1964) $Q$ 
parameter, which for gaseous disks must be less than unity for local 
gravitational instability.  Thus, also 
for later reference, we note that associated with equation (\ref{implicit_a2}) 
is a Toomre $Q$, which is given by the formula
$Q = \Omega a/\pi G\Sigma$ for a disk in quasi-Keplerian rotation.  In 
principle, $a$ should be computed from the considerations of heat balance 
outlined earlier, but for fiducial purposes, we use the value of $a$ associated 
with
case (\ref{A2}):
\be
Q = {3\over \sqrt{2I_\ell}} D \left[ A(1-f^2)\right]^{3/2} 
\left[ {M_*\over \dot M_{\rm d}}\left({GM_*\over \varpi^3}\right)^{1/2}\right].
\label{Qmin}
\ee
In what follows, we take the combination, $\lambda > 1$ and $Q < 1$, as 
necessary 
indicators for local gravitational instability.  With a ``standard'' flaring 
law, $A \propto \varpi^{1/4}$, both criteria favor the outer regions of a disk 
for the possible occurrence of disk fragmentation.

Finally, for arbitrary values of $f$, equations 
(\ref{viscousSig}) and (\ref{Prandtl}) give the viscosity and resistivity generated 
by the MRI instability in the disk as
\be
\nu=\frac{AD(1-f^2)}{f I_\ell} (GM_*\varpi)^{1/2},
\label{nu_pd}
\ee
\be
\eta=\frac{3A^2D(1-f^2)}{2f I_\ell^2}(GM_*\varpi)^{1/2}.
\label{eta_pd}
\ee
Apart from the factors involving $A$, $D$, and $f$, these expressions show that 
the natural scaling for both 
diffusivities is the specific angular momentum of the matter in Kepler orbits, 
$(GM_*\varpi)^{1/2}$. Notice especially the lack of any parametric dependence on 
the assumed mass accretion rate $\dot M_{\rm d}$. 

\subsection{Enclosed disk mass, magnetic flux, and angular momentum}

To make equations (\ref{bz_pd}) and (\ref{sig_pd}) consistent with equations 
(\ref{vertfield}) and (\ref{powerlawS}),
the product $DA$ has to be a power law of $\varpi$.  We adopt the natural 
assumptions that $D$ equals a 
constant and the disk flares as a power-law, so that the aspect ratio $A(r)$ at 
radius $r$ is related to its value $A(\varpi)$ at radius $\varpi$ by the 
formula:
\be
A(r) = A(\varpi)(r/\varpi)^n,
\label{diskflare}
\ee
where $n$ is the flaring exponent, which is positive definite if shadowing does 
not occur.  Had we adopted these assumptions from the start, together with the
hypothesis (\ref{def_nu}), we could have shown that not only are the discovered 
power-law solutions
possible in steady state, but they are unique. 
The relationship between the exponent $n$ and the exponent $\ell$ in equation 
(\ref{powerlawS}) can be found from equation 
(\ref{sig_pd}) with $A(\varpi) \propto \varpi^{n}$, namely, $-2\ell=-n-1/2$, or
\be
\ell = (1+2n)/4.
\label{elln}
\ee
For later reference we note from Table 1 that 
$I_\ell = 1.573$, $1.742$, or $2.188$ for disks with low, typical, or high power-law flaring, $n = 1/8$ and $\ell = 5/16$ (i.e., $\Sigma \propto \varpi^{-5/8}$), $n = 1/4$ and $\ell = 3/8$ (i.e., $\Sigma \propto \varpi^{-3/4}$), or 
$n = 1/2$ and $\ell = 1/2$ (i.e., $\Sigma \propto \varpi^{-1}$), respectively.

We wish now to compute the enclosed mass in the disk inside a radius $\varpi$,
\be
M_d(\varpi)\equiv 2\pi\int_0^{\varpi} \Sigma(r)r\,dr =
\left[\frac{4I_\ell}{3(3-2n)}\right]\left(\frac{f}{1-f^2}\right)
\left[\frac{1}{ D A(\varpi)}\right]
\frac{\dot M_{\rm d}}{(GM_*/\varpi^3)^{1/2}}.
\label{diskmass}
\ee
Thus, $M_d(\varpi)$ is a multiple of the mass that accretes 
through the disk during a Keplerian rotation period at that radius. This 
multiple
depends on the combination $(DA)^{-1}$ and on the effective value of $f$. 
Except for a numerical factor of order unity, we easily verify the 
interpretation that the enclosed mass $M_d(\varpi)$ results from accretion at a 
rate $\dot M_{\rm d}$ over a viscous diffusion time scale $\varpi^2/\nu$.

According to the discussion in \S 1.2, the magnetic field brought in by 
infall is contained as magnetic flux threading the disk. Inside any radius $\varpi$ where the disk is in steady state 
the enclosed magnetic flux is given by
\be
\Phi_d(\varpi) \equiv 2\pi\int_0^\varpi B_z(r)r \, dr =
2\pi\left(\frac{4}{3-2n}\right)\left[\frac{2f}{3DA(\varpi)}\right]^{1/2}(GM_*
\dot M_{\rm d}^2\varpi^3)^{1/4}.
\label{diskflux}
\ee
If the disk's mass is negligible in comparison with the star's, the system's
dimensionless mass-to-flux ratio enclosed inside $\varpi$ equals
\be
\lambda_*(\varpi) \equiv \frac{2\pi G^{1/2}M_*}{\Phi_d(\varpi)}=\left(\frac{3-2n}{4}\right)
\left[\frac{3DA(\varpi)}{2f}\right]^{1/2}\left(\frac{GM_*^3}{\dot 
M_{\rm d}^2\varpi^3}\right)^{1/4}.
\label{sysmasstoflux}
\ee

Infall models with field freezing until small radii yield $\lambda_* \approx 1$ 
to 4 (Galli et al.~2006),
consistent with the polarization findings of Girart et al. (2006) of an 
hourglass shape in NGC 1333 IRAS 4A. 
Field slippage during the collapse reduces the enclosed flux for a low-mass 
protostar
by a further factor of 2 to 3 at the radii $\sim 300$ AU that their disks are 
likely to occupy (see Fig.
3 of Shu et al. [2006] when $R_{\rm Ohm}$ in that paper has a value $\sim 10$ 
AU).  Thus, in \S 3, we shall adopt an enclosed mass-to-flux value of $\lambda_0 =4$ 
for the system
as a typical outcome of the star and disk formation process.
If insufficient time has elapsed for the disk to reach steady state inside the 
radius where $\lambda_* = \lambda_0$, 
investigations of the affected regions should make use of the time-dependent 
equations with which we began this paper (\S 1.1).

It is extremely informative to compute the enclosed mass at a radius $R_\Phi$ 
where $\lambda_*(R_\Phi) = \lambda_0$, i.e., at a radius where the disk contains 
the 
entire flux brought in by star formation:
\be
M_d (R_\Phi) =\left[ {(3-2n)I_\ell \over 8\lambda_0^2}\right]{M_*\over 1-f^2}.
\label{amazing}
\ee
For the typical case, $n = 1/4, \ell = 3/8$, and $I_\ell = 
1.742$, equation (\ref{amazing}) becomes
\be
1-f^2 = \left({0.5444\over \lambda_0^2}\right) {M_*\over M_d(R_\Phi)}.
\label{usefulforf}
\ee  
For a closed system in which infall has ceased, so that 
$\lambda_0$ remains a fixed constant, disk accretion must decrease $M_d(R_\Phi)$ 
relative to $M_*$, and therefore the departure from Keplerian rotation, 
$(1-f^2)$, must grow with time. This trend arises because viscosity drains mass 
from the disk onto the star, while resistivity can only cause the redistibution 
of flux within the disk but cannot change the total, making the specific 
magnetization (inverse $\lambda$) rise with time.  In \S 3 we shall combine
equation (\ref{usefulforf}) with an assumption of a disk's ``age'' to estimate    
the numerical value of the important parameter $f$.

To appreciate concisely the dynamical consequences of disk magnetization,
we note that equations (\ref{masstoflux}) and (\ref{sysmasstoflux}) imply the 
interesting reciprocity relationship:
\be
\lambda (\varpi)\lambda_*(\varpi) = \left({3-2n\over 4}\right)\left( 
{I_\ell\over 1-f^2}\right),
\label{reciprocity}
\ee
where the right-hand side is a constant that depends only on $f$ and $\ell = 
(1+2n)/4$.  Except for such constants,
a similar relationship was derived by Shu, Li, \& Allen (2004) in their 
analysis of magnetic levitation of pseudo-disks by strongly magnetized central 
objects (see their eq.~[65]).  Although disks differ from pseudo-disks in being 
(partially) centrifugally supported, and although the inner parts of a 
magnetized disk differ from a split monopole in their detailed interaction with 
the outer parts of the same magnetized disk, the principles are qualitatively 
similar and provide physical insight into why such magnetic support (levitation 
of the outside by the inside) causes the rotation to occur at sub-Keplerian 
rates.  

Finally, if the mass is
mostly in $M_*$, the enclosed angular momentum of the parts of the disk that are 
in steady-state accretion is given by
\be
{\cal J}_d =2\pi\int_0^{\varpi} \Sigma(r) r^2\Omega (r) r\, dr
=f\left(\frac{3-2n}{4-2n}\right)M_d(\varpi) (GM_* \varpi)^{1/2}.
\label{sysangmom}
\ee
Equations ({\ref{diskflux}), (\ref{diskmass}), 
and (\ref{sysangmom}) show that the enclosed disk flux, mass, and angular 
momentum represent a sequence of decreasing central concentration.

\subsection{Semi-numerical formulae}

For the convenience of the reader, we express the results of the analytical 
theory in the following semi-numerical form (assuming $n=1/4$):
\begin{eqnarray}
B_z(\varpi) & = & 8.89 \times 10^{-3}\,
D^{-1/2} 
\left(\frac{M_*}{0.5\, M_\odot}\right)^{1/4}   
   \left(\frac{\dot M_{\rm d}}{2 \times 10^{-6} M_\odot\, {\rm yr}^{-1}}\right)^{1/2} 
\nonumber \\
 & &  \times f^{1/2}\left[{0.1 \over A(\varpi)}\right]^{1/2} 
\left(\frac{\varpi}{100 \, {\rm AU}}\right)^{-5/4} \, {\rm G},
\label{num_Bz}
\end{eqnarray}
\begin{eqnarray}
\Sigma(\varpi) & = &  0.740\,
D^{-1}
\left(\frac{M_*}{0.5 \, M_\odot}\right)^{-1/2}
   \left(\frac{\dot M_{\rm d}}{2 \times 10^{-6} M_\odot \, {\rm yr}^{-1}}\right) 
\nonumber 
\\
 & & \times \left( \frac{f}{1 - f^2} \right)
 \left[\frac{0.1}{A(\varpi)}\right]\left(\frac{\varpi} {100 \, {\rm 
AU}}\right)^{-1/2} \, {\rm g \, cm^{-2}},
\label{num_Sigma}
\end{eqnarray}
\begin{eqnarray}
 M_d(\varpi) & = & 4.18 \times 10^{-3}\, 
 D^{-1}
 \left(\frac{M_*}{0.5\, M_\odot}\right)^{-1/2}
    \left(\frac{\dot M_{\rm d}}{2 \times 10^{-6}\, M_\odot \, {\rm yr}^{-1}}\right) 
\nonumber \\
 & & \times \left( \frac{f}{1 - f^2} \right)
\left[\frac{0.1}{A(\varpi)}\right] \left(\frac{\varpi}{100 \, {\rm 
AU}}\right)^{3/2}\, M_\odot ,
\label{num_Md}
\end{eqnarray}
\begin{eqnarray}
 {\cal J}_d(\varpi) & = & 0.629\, 
 D^{-1}
       \left(\frac{\dot M_{\rm d}}{2 \times 10^{-6} M_\odot \, {\rm yr}^{-1}}\right) 
\nonumber \\
& & \times \left( \frac{f^2}{1 - f^2} \right)
\left[\frac{0.1}{A(\varpi)}\right] \left(\frac{\varpi}{100 \, {\rm 
AU}}\right)^{2} M_\odot  \, {\rm km \, s^{-1}}\, {\rm AU},
\label{num_Jd}
\end{eqnarray}
\begin{eqnarray}
\nu(\varpi) & = & 1.81 \times 10^{19}\,
D
 \left(\frac{M_*}{0.5\, M_\odot}\right)^{1/2}  \nonumber \\
 &  & \times \left( \frac{1 - f^2}{f} \right)
 \left[ {A(\varpi)\over 0.1}\right] \left(\frac{\varpi}{100 \, {\rm 
AU}}\right)^{1/2} \, {\rm cm^2 \, s^{-1}},
 \label{num_nu}
\end{eqnarray}
\begin{eqnarray}
\eta(\varpi) & = & 1.56 \times 10^{18} \,
D
 \left(\frac{M_*}{0.5 \, M_\odot}\right)^{1/2}  \nonumber  \\
  &  & \times \left( \frac{1 - f^2}{f} \right)
 \left[{A(\varpi})\over 0.1\right]^2 \left(\frac{\varpi}{100 \, {\rm 
AU}}\right)^{1/2} \, {\rm cm^2 \, s^{-1},}
 \label{num_eta}
\end{eqnarray}
\begin{eqnarray}
\lambda(\varpi) & = & 0.135\,
\, D^{-1/2}
 \left(\frac{M_*}{0.5\, M_\odot}\right)^{-3/4} 
 \left(\frac{\dot M_{\rm d}}{2 \times 10^{-6} M_\odot\, {\rm yr}^{-1}}\right)^{1/2} 
\nonumber  \\
   &  & \times \left( \frac{f^{1/2}}{1 - f^2} \right)
\left[ \frac{0.1}{A(\varpi)}\right]^{1/2} \left(\frac{\varpi}{100\, {\rm 
AU}}\right)^{3/4},
\label{num_lambda}
\end{eqnarray}
\begin{eqnarray}
\lambda_*(\varpi) & = & 8.07\,
D^{1/2}
 \left(\frac{M_*}{0.5 M_\odot}\right)^{3/4} 
  \left(\frac{\dot M_{\rm d}}{2 \times 10^{-6} M_\odot \, {\rm yr}^{-1}}\right)^{-1/2} 
\nonumber  \\
     &  & \times 
     f^{-1/2}\left[\frac{A(\varpi)}{0.1}\right]^{1/2} \left(\frac{\varpi}{100\, 
{\rm AU}}\right)^{-3/4},
     \label{num_lambda*}
\end{eqnarray}
\begin{eqnarray}
Q & = & 56.4\,  D \,
 \left({M_*\over 0.5 \, M_\odot}\right)^{3/2}\left( {\dot M_{\rm d}\over 2\times 
10^{-6} M_\odot \, {\rm yr}^{-1}}\right)^{-1}
\nonumber \\
   & & \times
 \left[ {(1-f^2)^3\over f}\right]^{1/2}\left[{A(\varpi)\over 0.1}\right]^{3/2} 
\left({\varpi \over 100 \, {\rm AU}}\right)^{-3/2}.
\label{num_Qmin}
\end{eqnarray}

\section{ASTRONOMICAL EXAMPLES FROM STAR FORMATION}

To give astronomical context to the theory developed so far, we consider four 
examples of interest in current-day star formation: (1) a T Tauri star, (2) an 
embedded low-mass protostar,
(3) an FU Orionis star, and (4) an embedded high-mass protostar.
Models 1, 2, 3 have a central star of mass 
0.5 $M_\odot$, and differ only in accreting at rates
equal, respectively, to $1\times 10^{-8}$ $M_\odot$ yr$^{-1}$ (Gullbring et al. 
1998), $2\times 10^{-6}$ $M_\odot$ yr$^{-1}$ (Young et al. 2003, Young \& Evans 
2005), and $2\times 10^{-4}$ $M_\odot$ yr$^{-1}$ (see Popham et al. 1996, who, 
however, have a different explanation
for sub-Keplerian rotation near the star than this paper).
Model 4, the high-mass protostar, has a mass-accretion rate that is scaled 
relative to Model 2, the low-mass protostar, by their
masses (see, e.g., Stauber et al. 2007), i.e., both $M_*$ and $\dot M_{\rm d}$ are 
taken to be a factor of 50 larger.  In each YSO disk, we assume a standard 
flaring law 
(see, e.g., the dashed curve in Fig. 1b of D'Alessio et al. 1999),
\be
A(\varpi) = 0.1 (\varpi/100\; {\rm AU})^{1/4}.
\label{adoptedflaring}
\ee 

We assign $t_{\rm age}$ = $3\times 10^6$ yr (Haisch, Lada, \& Lada 2001), 
$1\times 10^5$ yr (Jijina, Myers, \& Adams 1999), $100$ yr (Herbig 1977), and 
$1\times 10^5$ yr (Osorio, Lizano, \& D'Allesio 1999) as the fiducial ages, 
respectively, of the T Tauri star, low-mass protostar, FU Orionis outburst, and 
the high-mass protostar.
We now compute a viscous-accretion radius $R_\nu$ such that $M_d(R_\nu)/\dot M_{\rm d} 
= t_{\rm age}$.  To ensure the approximate validity of the steady-state 
assumption, we then set $R_\nu = R_\Phi$, where $R_\Phi$ is defined as before to 
equal the radius that contains all the flux, i.e., 
$\lambda_*(R_\Phi) = \lambda_0$.  Since $M_d(R_\nu) = M_d(R_\Phi)=\dot M_{\rm d} 
t_{\rm age}$ in this formalism, the departure from Keplerian rotation can be 
computed from equation (\ref{usefulforf}) to equal
\be
1-f^2 = {0.5444 M_* \over \lambda_0^2 \dot M_{\rm d} t_{\rm age}}.
\label{selfconsistentf}
\ee
For protostars still building up their mass, we expect $\dot M_{\rm d} t_{\rm age}$ to 
be comparable to $M_*$.  On the other hand, for T Tauri stars or FU Orionis 
objects, we have $\dot M_{\rm d} t_{\rm age}$ small compared to $M_*$.  Thus, for 
$\lambda_0$ of order 4, we anticipate the departures 
from Keplerian rotation to be more substantial for
T Tauri and FU Orionis stars than for low- or high-mass protostars.

The surface density $\Sigma$ must drop faster with $\varpi$ than any negative 
power law in the outer parts of the disk in order to vanish, by definition, at 
some true outer disk edge $R_{\rm d}$. Therefore, unlike the assumption being made at 
$R_\nu$, the term 
represented by the right-hand side of equation (1-3) must change sign in the 
outermost parts of the disk, leading to a viscous movement outward of $R_{\rm d}$ with 
time, carrying to large distances much of the system's angular momentum. Hence, 
the enclosed angular momentum, calculated from equation (\ref{sysangmom}) with 
$\varpi = R_\Phi$, may not yield a representative estimate of the system's true 
total store of angular momentum because it is the least centrally concentrated 
of the trio: magnetic flux, mass, and angular momentum.

The numerical values of the relevant input parameters are now tabulated in 
columns 2 through 5 of Table 2; the output bulk parameters from the theory are 
tabulated in columns 6 through 11. In accordance with the discussion of \S 2.3 
we have chosen $\lambda_0 = 4$ for all four cases.  This choice results in $f = 
0.957$ for the low-mass and high-mass protostar models, making 
magnetocentrifugally-driven, cold, disk-winds, unassisted by photo-evaporation, difficult but not impossible at radii much less than 100 AU for these systems, according to the criterion (\ref{diskwindcrit}) for appreciable disk 
winds.  If we had chosen a smaller value, say 2, for $\lambda_0$, then $f$ would have 
equaled $0.812$, and a powerful disk wind in protostars at any radii other than the inner disk-edge, 
where magnetospheric interactions dominate, would have been almost 
as unlikely as the $f = 0.658$ and $f = 0.386$ cases tabulated for the T Tauri stars and FU Orionis systems 
with $\lambda_0 = 4$.

\begin{deluxetable}{lllllllllll}
\rotate
%\tabletypesize{\scriptsize}
%\setlength{\tabcolsep}{0.04in}
\tablecolumns{11}
\tablewidth{0pc}
\tablecaption{Parameters of Four Model Systems}
\tablehead{
\colhead{Object} & \colhead{$M_*$} & \colhead{$\dot M_{\rm d}$} & 
\colhead{$t_{\rm age}$} & \colhead{$D$} & \colhead{$f$} & \colhead{$R_\Phi=R_\nu$} 

& \colhead{$M_d(R_\Phi)$} & \colhead{${\cal J}_d(R_\Phi)$} & 
\colhead{$\lambda(R_\Phi)$} & \colhead{$Q(R_\Phi)$} \\
\colhead{} & \colhead{($M_\odot$)} & \colhead{($M_\odot$/yr)} & \colhead{(yr)} & 

\colhead{} & \colhead{} & \colhead{(AU)} & \colhead{($M_\odot$)} & 
\colhead{($M_\odot$~AU~km/s)} & \colhead{} & \colhead{} \\
}
\startdata

T Tauri star        & 0.5 & $1\times 10^{-8}$ & $3\times 10^6$ & $10^{-2.5}$ & 
0.658 & 298   & 0.0300 & 5.12   & 0.480 & 4.47  \\
Low-mass Protostar  & 0.5 & $2\times 10^{-6}$ & $1\times 10^5$ & 1           & 
0.957 & 318   & 0.200  & 51.4   & 3.20  & 0.381 \\
FU Ori              & 0.5 & $2\times 10^{-4}$ & $1\times 10^2$ & 1           & 
0.386 & 16.5  & 0.0200 & 0.473  & 0.320 & 3.36  \\
High-mass Protostar & 25  & $1\times 10^{-4}$ & $1\times 10^5$ & 1           & 
0.957 & 1,520 & 10.0   & 39,700 & 3.20  & 0.463 \\

\enddata
\end{deluxetable}

\subsection{Discussion of Bulk Properties}

In Table 2 we have chosen the numerical value of $D$ to make $R_\Phi = R_\nu 
\propto D^{4/5}$ reasonable. It is informative in this regard that $D = 1$ works 
for the two protostar and FU Orionis models (perhaps $D = 0.3$ would be better), 
but only a relatively small value, $D = 10^{-2.5}$, does as well for the T Tauri 
model (see, e.g., Andrews \& Williams 2007).  The enclosed disk mass $M_d(R_\Phi)$ is independent of the numerical 
value of $D$, whereas the enclosed angular momentum ${\cal J}_d(R_\Phi)$ 
varies as $D^{2/5}$.   

In the case of the FU Orionis model, we are contemplating a {\it transient} 
accretion event that has occurred during the past 
100 yr, that 
has swept most of the mass and the magnetic flux to within a radius $\sim R_\Phi 
= 16.2$  AU, inside of which the system is in quasi-steady outburst. In this 
case, $R_\nu = R_\Phi$ is likely to be considerably smaller that $R_{\rm d}$, where 
most of the system angular momentum may still reside.
In the other cases, we think of $R_\Phi = R_\nu$ as being the effective ``outer 
radius'' of the system, large enough to contain the magnetic flux that was 
dragged into the system by the star formation process, but  small enough so that 
the available viscosity is able to establish a quasi-steady state if the angular 
momentum contained in the system is comparable to that tabulated as ${\cal 
J}_d(R_\Phi$).  

Near the disk edge $R_{\rm d}$ where $\Sigma$ becomes vanishingly small
and $B_z$ matches onto interstellar values, 
radial magnetic-buoyancy effects with interchange and/or Parker instabilities 
may lead to a net loss of flux from the disk. Continuing infall that brings in 
additional mass and flux from the cloud-core surroundings may counteract such 
tendencies. 
Moreover, equations (\ref{diskmass}) and (\ref{diskflux}) show that the outer 
parts of disks contain relatively more mass than flux, so stripping of the outer 
parts by edge effects, or by tidal encounters with other stars, or by 
photo-evaporation from the far ultraviolet produced by the most massive cluster 
members in dense clusters (Adams et al. 2006), 
cannot do much to alleviate
the problem of the growing magnetization of the entire disk. In what follows, we 
ignore the complications that may arise from all such environmental effects.

One could try to justify very small values of $D$ = $10^{-3}$ to $10^{-2}$ in T 
Tauri models on the basis that only the top and bottom 10 g cm$^{-2}$ of a disk 
(from ionization by scattered X-rays; see Igea \& Glassgold 1999), containing a 
total $\Sigma$ of 2000 or 200 g cm$^{-2}$ (that apply roughly at 1 AU and 5 AU 
of the ``minimum solar nebula''), are active in the accretion process (Gammie 
1996).  Such small values for the effective $D$ are not out of the question if T 
Tauri disks have substantial ``dead zones'' where the ionization is too low to 
couple to magnetic fields except, possibly, for thin surface layers.  The 
surface density ratio of active zone to dead zone need not be as small as 
$10^{-2.5}$ (see \S 3.2) to result in such a value for $D$, if accretion in a 
thin surface layer has an intrinsically smaller efficiency than fully turbulent 
MRI (see the discussion in \S 4.2).

A related problem arises when we apply equation (\ref{implicit_a2}) to our T 
Tauri model, which results in the expression, $a \approx 2.64 
(\varpi$/AU)$^{-3/8}$ km s$^{-1}$.  An isothermal sound speed of 2.64 km 
s$^{-1}$ corresponds to a temperature in molecular gas of $\sim 1,900$ K, a 
value that is unlikely to hold even in the mid-plane at 1 AU.  The difficulty 
arises because we took the first term $\propto A$ in equation 
(\ref{implicit_a2}) to be dominant, which requires vigorous inward accretion to 
sustain the assumed $B_\varpi^+/B_z = I_\ell = 1.742$ that accounts for the 
substantial departure from Keplerian rotation $f = 0.658$ computed in Table 2, 
yet we took the diffusion constant $D$ to equal a paltry $10^{-2.5}$. The 
inconsistency disappears if we assume that the inner disks of T Tauri stars are 
dead to the MRI except for their superficial layers.  With no magnetic coupling 
in the deeper layers, $f$ would be much closer to unity in the midplane than 
indicated in Table 2, and the second term $\propto A^2$ in equation (\ref{implicit_a2}) would 
dominate.  We then easily compute that midplane temperatures at 1 AU would be 
closer to 600 K, and dropping as $(\varpi/{\rm AU})^{-1/2}$, more in line with 
conventional estimates.  However, we end with a non-standard picture where the 
central dead layers of T Tauri disks rotate at near-Keplerian speeds, while the 
active superficial layers tend to be very sub-Keplerian.  This picture raises 
speculative conjectures that we defer to \S 4.2.

The enclosed mass $M_d(R_\Phi)$ for the T Tauri and FU Orionis models, which are 
independent of the choice of effective $D$ (as long as it is a constant as a 
function of $\varpi$) are similar to standard estimates. The disk mass of the 
low-mass protostar is comparable to those found by Jorgensen et al.~(2007 and 
references therein) and exceed a value
equal to the ``maximum solar nebula'' when self-gravitational instabilities 
would need to be considered (Shu, Tremaine, Adams, \& Ruden 1990).  The disk 
masses of high-mass protostars are not observationally well studied because of 
their rarity and consequent larger distances. If the high disk masses for the 
two protostar models were real, gravitational instabilities could lead to disk 
fragmentation because $\lambda (R_\Phi)$ and $Q(R_\Phi)$ are, respectively, 
greater and less than unity in their outer disks.  

In contrast, the FU Orionis and T Tauri models are stable to disk fragmentation 
by both the magnetic and Toomre criteria, $\lambda < 1$ and $Q > 1$. While these 
conclusions do depend somewhat on the specific choices made for $D$, the 
formation of either gaseous giant planets or brown dwarfs by gravitational 
instability at tens of AU or smaller can probably be ruled out in the model T 
Tauri system. Thermal cooling is not sufficiently rapid to help (Rafikov 2007).
Certainly, close-in brown-dwarf companions that could have been 
easily detected by the Doppler method seem difficult to produce in all our 
models, which is an after-the-fact explanation for the so-called ``brown-dwarf 
desert'' (Marcy \& Butler 2000, Halwachs et al. 2000).  Our speculations in this 
regard are consistent with the intuitive notion that large-angular momentum 
cases are more prone to making binaries by gravitational fragmentation, whereas 
small angular-momentum cases are more prone to making planetary systems by the 
embryonic core-accumulation of solids (Lin 2006).

\subsection{Disk surface densities and magnetic fields}

Figure 2 shows $\Sigma$ (in units of g cm$^{-2}$) and $B_z$ (in units of G), 
computed from equations (\ref{sig_pd}) and (\ref{bz_pd}), as functions of 
$\varpi$ (in units of AU) for the four model systems.  
Hexagons have been placed as stop signs on the formal plots to indicate that the 
curves for radii larger than $R_\Phi = R_\nu$  should be ignored in any 
realistic applications.  
Consistent with Table 2, we have chosen $D = 1$ for the two protostars and FU 
Orionis, and $D = 10^{-2.5}$ for the T Tauri model.   

In the low-mass protostar model, $B_z$ is 302 G, 1.09 G, and 8.74 mG at $\varpi$ 
= 0.05 AU, 3 AU, and 100 AU, respectively, for $D = 1$. In the high-mass 
protostar model, they are higher by a factor $50^{3/4} = 18.8$ at each radius. 
The first value in a low-mass protostar, 302 G at 0.05 AU, is compatible with a 
more-or-less smooth matching (after enhancement in the X-region) with the 
inferred field strength of the stellar magnetopause (see Shu et al. 1994), 
yielding yet another indication why disk truncation occurs inside such radii 
when the stellar field increases inward much more strongly than the extrapolated 
disk field.

The second value, 1.09 G at 3 AU, is compatible with the paleomagnetism measured 
for the chondrules of primitive meteorites, whose parent bodies are believed to 
originate in the asteroid belt 
(reviewed by Stacey 1976, Levy \& Sonett 1978, and Cisowski \& Hood 1991). 
In the X-wind model, whose predictions on this point have received strong 
support from the recent Stardust comet-sample mission (McKeegan 2006, Zolensky 
et al. 2006), the heating of the chondrule-like materials and refractory 
inclusions found in meteorites, and now comets, occurs close to the protosun 
(Shu, Shang, \& Lee 1996; Shu et al. 1997). Nevertheless, when ferromagnetic 
chondrules are thrown 
out to the asteroid belt, they should not encounter fields that are much 
stronger than the inferred paleomagnetism result of 1 to 10 G (Stacey 1976,
Levy \& Sonett 1978, and Cisowski \& Hood 1991).

The third value, 8.74 mG at 100 AU, offers an inviting target for Zeeman 
measurements if appropriate sources of maser emission can be found in 
protostellar disks.  Even more promising for such studies, because maser 
emission in ring-like configurations 
has already been found (Hutawarakorn \& Cohen 2005; Edris, Fuller, \& Cohen 2007), are the large disk 
fields predicted for high-mass protostars if their mass-accretion rates scale 
anything like their mass even at radii of a few hundred AU.

Consider now the FU Orionis model.  At $\varpi =$ 0.05 AU, not far from the 
putative stellar surface, the vertical magnetic field is predicted to be 1.92 kG 
if $D=1$, somewhat higher than the value for $B_z \sim$ 1 kG inferred from 
observations for the inner disk of FU Orionis itself (Donati et al. 2005).  A 
four-times thicker disk, as may happen for the hot inner regions, would 
eliminate the discrepancy.  Moreover, the same observations claim that the inner 
regions of the disk in FU Ori rotate at a speed 2 to 3 times lower than the 
Keplerian value, consistent with $f = 0.386$ in our model.  A more detailed 
study of this system is warranted, but we caution that precise modeling would 
need to take into account the interaction of the magnetized accretion disk with 
the (squashed) stellar magnetosphere.  If the mass accreted onto the star per FU 
Orionis event is $M_d(R_\nu) \sim 0.02 \, M_\odot$ independent of $D$, as 
modeled in Table 2, then it takes tens of such events to accumulate the entire mass of the star,
in accord with the astronomical statistics 
of such objects (Hartmann \& Kenyon 1996, but see also Herbig et al. 2003).    

Figure 2 shows that the surface densities for the T Tauri model disk are 42.9 
and 7.63 g cm$^{-2}$ at 1 and 10 AU, respectively, if $D = 10^{-2.5}$.  These 
are smaller values for the planet-formation zones of terrestrial and giant 
planets than given by conventional minimum solar nebulas (Hayashi et al. 1985) 
because a comparable disk mass is spread over a much larger area (but see \S 4.2).   
Indeed, $\Sigma$ at $\varpi > 100$ AU has dropped to such low values (below visual extinctions of
1 mag to the mid-plane) that any CO would be dissociated by the interstellar radiation field.  This
phenomenon and the natural steepening of $\Sigma$ near the outer edges of accretion disks mentioned earlier may account
for some of the larger power-law gradients inferred from CO brightnesses reviewed by Dutrey, Guilloteau, \& Ho~(2007).
Notice that the inferred magnetic field strength at 3 AU is
1.13 G, which is compatible with the chondritic values, and indeed has not 
changed much from the case of the low-mass protostar. The near-equality arises 
because the low-mass protostar and T Tauri models have about the same flux and 
the same radius, although their disk masses differ by a factor of 6.67 and their 
angular momenta by a factor of 10 (for the chosen values of $D$). Thus, Figure 2 
shows that the surface densities differ by a factor of a little over 6, but the 
two $B_z$ curves lie almost on top of one another all the way out to about the 
same position for the two hexagons.  This result makes graphic the point that 
the T Tauri disk rotates slower than the low-mass protostar disk because the 
former is more strongly magnetized relative to the disk mass. 
Indeed, one could heuristically imagine the low-mass protostar evolving into the 
T Tauri system if the excess mass and excess angular momentum could be put into 
an orbiting stellar companion without changing the magnetic flux distribution.

With $D$ equal to a strict constant, the surface densities  
corresponding to low, typical, and high power-law flaring, $A(\varpi)\propto \varpi^{1/8}$, $\varpi^{1/4}$, and $\varpi^{1/2}$ are, respectively, $\Sigma \propto \varpi^{-5/8}$, $\varpi^{-3/4}$, and $\varpi^{-1}$, consistent with the power-law range deduced from the thermal dust-emission of YSO disks (Andrews \& Williams 2007), but shallower than the law $\Sigma \propto \varpi^{-3/2}$ associated with conventional minimum solar nebulas (Hayashi et 
al.~1985).  However, for T Tauri stars, once one allows the possibility that $D$ might be 
substantially smaller than unity because of physical considerations other than 
fully developed MRI turbulence (see below), then there is no reason to think 
that $D$ would be a spatial constant.  On the other hand, we may do well to 
recall that the steeper, empirical, log-log slope is derived from the inferred 
distribution of solids, which may, as seems to be implied by the Comet Wild 
results (McKeegan 2006, Zolensky et al. 2006), have been affected relative to 
the gas by the recycling of rock from the hot disk regions near the protosun to 
the rest of the solar nebula, as well as by the migration of planets.  In any case, we would be the first to admit that 
our models do not allow for a straight-forward recovery of models that look like 
the ``minimum solar nebula.''  Probably no viscous accretion disk can ``succeed''
in this regard (see Fromang, Terquem, \& Balbus 
2002).

Vorobyov \& Basu (2006) suggest that FU Orionis outbursts are associated with 
spiral gravitational instabilities in a protostellar disk.  We are sympathetic 
to the view that such self-gravitational disturbances can play a role in the 
early evolution of protostars that are still in the main infall stage that 
builds up the final system mass (Shu et al. 1990).  We are however agnostic when 
it comes to the issue whether FU Orionis systems represent such early-stage 
objects or not. Accurate estimates of the outburst disk masses in FU Orionis 
systems -- whether they are closer to ``minimum'' or ``maximum'' values -- can 
prove to be observationally decisive in this debate.

\section{HIGH AND LOW STATES OF ACCRETION}

In \S 3 we have presented the astrophysical case that there are high states of 
accretion where $D$ is of order unity (FU Ori, low- and high-mass protostars) 
and low states where $D$ is much less than unity (T Tauri stars).  Indeed, even 
low-mass protostars (or, at least, the so-called Class I sources) may alternate 
between high- and low-states of mass accretion (White et al. 2007).  In \S 4.1 
that follows, we speculate that the MRI is fully developed in high states, and 
we consider mechanisms that may allow turbulent values of $\nu$ and $\eta$, in a 
situation with dynamically strong mean fields, to achieve the saturated ratio of 
equation (\ref{Prandtl}) in steady state.  Likewise, in \S 4.2 that follows, we 
discuss the weak-turbulence conditions likely to prevail on scales smaller than 
a vertical scale height $z_0$ in low states, focusing in particular on the form 
of MRI likely to be present when dead zones are bottlenecks to rapid disk 
accretion, with the activity being concentrated in thin surface layers (Gammie 1996).

\subsection{Magnetic loops and their dynamics}

To picture how rapid transport of matter and magnetic fluctuations across strong 
mean field lines 
that are anchored externally is possible, we adopt a mental image of field loops 
on a scale smaller than $z_0$.
This mental image can be given a physical correspondence in the equations of 
magnetohydrodynamics in axial symmetry, but the task becomes much harder if the 
current associated with the loop has structure in the $\varphi$ direction.  We 
ignore this
complication in the heuristic discussion based on a diagram (Fig. 3) that shows 
only one field loop, born of a single
mean field line, that has complete freedom to move as if there were no 
constraints from neighboring field lines and other loops.
Because we make no attempt to be quantitative except for a single 
order-of-magnitude calculation, this mental image can suggest
possible interpretations without misrepresenting, hopefully, the complex, 
nonlinear dynamics of fully developed, 3-D MRI turbulence. 

Consider a process that bends, pinches, and twists a field line 
into a loop that eventually disconnects from its parent field line by resistive 
dissipation (bottom set of diagrams in Fig. 3).
The loop is then advected to the next set of field lines, to which it 
reconnects, relaxes, and gets into position to form another loop.  To visualize 
what is happening in this figure, recall that magnetic field lines never end, 
but are directed continuously from point to point on a given line, except during 
reconnection, when oppositely directed fields can annihilate, leaving the 
remaining fragments to join up in a new field-line configuration. In a random 
field of fluid turbulence with a straight and uniform distribution of 
the mean field, a loop is as likely to get transported away from the star as 
toward it; i.e., the loops do a random walk, and the entire process is 
describable as a ``diffusion'' across mean field lines.   
The process has directionality and becomes a diffusive flux when the mean field 
has a spatial curl defined by the mean accretion flow, as drawn in the diagram.

Note that the entire``bend-pinch-disconnect and twist-reconnect'' process 
requires helical turbulence in three dimensions, with differential rotation 
providing the critical "twist" part of the process. Similar
diagrams were drawn by Parker (1955) in his famous proposal for the mechanism of 
dynamo action.  In the present
context, the ``twist" also implies an outward transport of angular momentum if 
it is accompanied by the shearing of the radial field $B_\varpi$ to give an 
azimuthal component $B_\varphi$.  In this manner, the entire sequence of steps 
provides both an effective viscosity for angular momentum transport and an 
effective resistivity for matter to move from one set of field lines to another.
In 2-D, ``bend-pinch-disconnect" gives a loop that can transport angular 
momentum if the shear of differential rotation generates a $B_\varphi$ from the 
local $B_\varpi$ (a ``2.5 D'' process).  But the loop, without an additional 
vortical ``twist'' in the third direction (requiring an ``eddy'' motion out of 
the page, azimuthally in the drawing), has the wrong orientation to attach to 
the next set of mean field lines (top set of diagrams in Fig. 3 with the fourth 
"reconnect" step forbidden by the large red cross).  Thus, the loop will be 
trapped between the thicket of nonzero mean-field lines, and it will eventually 
retrace steps 3-2-1 and merge back onto the original field line (or with other 
loops carrying the same sense of current), causing the matter to become 
re-attached more-or-less to the same field location except for the slight 
diffusion associated with dissipation by the microscopic collisional 
resistivity.  In such a situation, the fluctuations associated with MHD
turbulence are probably better described as a random collection of Alfv\'en waves 
rather than as a diffusing, merging, set of magnetic loops.   

The 3-D process appears in many MRI simulations when the plasma beta is large 
compared to unity ($\sim 100$, see Appendix B).  It remains to be seen if it 
persists in the presence of a mean field as dynamically strong as we advocate in 
this paper. In any case, the random walk of magnetic loops, carrying an 
associated current, can move through the thicket of mean field lines faster than 
individual particles or particulates can get knocked off one set of field lines 
by physical collisions to attach onto the next set of field lines, allowing for 
``viscous'' and ``resistive'' diffusivities that are larger than conventional 
microscopic values. The magnetic dissipation process is sometimes described by 
{\it hyper-resistivity}, i.e. turbulent transport of current, not field, which 
was originally proposed to describe magnetic relaxation in plasmas (Strauss 
1976, Diamond \& Malkov 2003). 
Appendix A shows how the simplest mathematical model of a diffusion of 
$\varphi$-current rather than a diffusion 
of $z$-field yields the same practical results as \S 2, but with a Prandtl ratio 
$\eta_J/\nu \sim 1$ rather than 
$\eta/\nu \ll 1$. 
 
In a pure-loop picture, the derivation of \S 2.1 really applies then to the loop 
dynamics of Figure 3. In other words, $B_\varpi$ of that derivation is really 
$\delta B_\varpi$ of the loop, with the change in sign of $\delta B_\varpi$ from 
the top to the bottom of the loop being irrelevant because the $\delta B_\phi$ 
that is produced by shearing will have the correct compensating sign as already 
noted in the discussion of \S 2.1. We then assume $\delta B_\varpi \sim 
B_\varpi^+$ and $\delta u \delta B_\varphi \sim \varpi (d\Omega/d\varpi)  \delta 
\varpi \delta B_\varpi$, with the corresponding Maxwell stress calculated from 
the quadratic correlation 
of $\delta B_\varpi \delta B_\varphi$ assuming $\delta u \sim \Omega \delta 
\varpi$ as in \S 2.1. The estimate for $\nu$ then goes through as before, with 
the large uncertainties in proportionality constants absorbed in ${\cal F}$ and 
eventually $D$. Consistent with the discussion of the formulation of the 
mean-field MHD equations in \S 1.1, we then have a mathematical separation in 
which there is no mean $B_\varphi$ when averaged over $z$, but there are local 
fluctuations $\delta B_\varphi$ whose correlations with $\delta B_\varpi$ do not 
average to zero.

The estimate for $\eta \sim A \nu$ with $A = z_0/\varpi \ll 1$ might follow 
because all detached and non-detached loops can transport angular momentum but 
only a fraction of the detached ones have the right geometry and orientation to 
reconnect with mean downstream field lines, yielding an effective resistivity 
$\eta$ that is much smaller than the turbulent viscosity $\nu$.   The exact 
relation (\ref{Prandtl}), which includes an extra factor of $3/2I_\ell$, then 
presumably arises because, in steady state, the rotation law is quasi-Keplerian 
and the surface density has a power-law index $-2\ell$ (eq.~[\ref{powerlawS}]). 
The fact that the scale of the turbulent mixing length $\delta\varpi$ was left 
unspecified in \S 2.1 (in actuality, a spectrum of such scales and shapes) may 
give the problem the necessary degree of freedom to make matters come out 
exactly right.  At some basic level, the macrophysics of fully developed MRI 
makes angular-momentum transport the driving energy-release mechanism behind the inward fluid drift in disk accretion.  The 
formation rate, merger rate, and geometry of magnetic loops may be regulated to 
yield a turbulent resistivity, $\eta = (3A/2I_\ell)\nu$, that is well below the 
naive Prandtl ratio $\eta \approx \nu$ because no energy source exists to cause 
gas to diffuse across flux tubes faster than the saturated value. Conversely, if 
$\eta$ were to fall below the level $(3A/2I_\ell)\nu$, the resultant pile-up of 
mean field lines waiting to diffuse inward (see LPP, whose solution for the 
induction equation remains valid independent of any implicit or explicit 
assumptions about $\Omega$) would presumably cause a shift in the numbers and 
kinds of loops generated until $\eta$ approaches the saturated level. However, 
more rigorous theoretical studies and/or numerical simulations are needed if we 
are to gain confidence that MRI dynamics under the circumstances envisaged in 
this paper can truly satisfy the diffusivity-ratio constraint implied by equation (\ref{Prandtl}) with $\nu$ given by equation (\ref{def_nu}).  If such studies show that equation (\ref{Prandtl}) cannot 
be achieved, then a possible resolution for real systems is to alternate between 
trying to satisfy $u=-3\nu/2\varpi$ and $u=-(\eta/z_0)B_\varpi^+/B_z$, making 
relaxation oscillations between the two conditions the real cause of low states 
and high states, with FU Orionis outbursts and their decay 
as the transition phenomenon.

In our enthusiasm for magnetic loops, we should not forget that Alfv\'en waves can 
also carry 
angular momentum, depositing it in the matter when they dissipate. If the MRI is 
operating at maximum efficiency, it is easy to show that the frequency 
associated with Alfv\'en waves of wavenumber scale $z_0^{-1}$ is comparable to 
$\Omega$.  For larger wavenumbers (smaller scales), the Alfv\'en wave frequency is 
larger than the natural eddy turn-over frequency $\Omega$, leading us to a 
picture of the generation of Alfv\'en waves by the bending or twisting of 
proto-loops of scale $z_0$  that do not detach from their mean field lines, 
before these wavelike disturbances propagate, go into a free cascade, and 
dissipate from interactions of the type described by Goldreich \& Sridhar's 
(1997) theory of MHD turbulence. It is unlikely that the competition with 
loop detachment and merging from such wave-transport effects could be adequately 
described by diffusion equations at a macro level.

\subsection{Two-dimensional MHD turbulence and layered accretion}

Creating loops of field and chopping them off from mean field lines by turbulent 
fluid motions may not be possible
when the coupling to magnetic fields is weak, and the MRI transport mechanism 
becomes confined to surface layers where the ionization level is still 
sufficiently high (the situation in T Tauri disks).  The formation of the loops 
themselves becomes difficult because the restricted height  practically 
available for $z$-motions may make the fluid effectively two-dimensional.  In 
particular, lifting parcels of gas against their own weight in the $z$-direction 
either to bend field lines or to twist them, added to the energy needed to 
stretch and pinch magnetic fields, may prove relatively difficult in thin 
surface layers compared to the same processes near the mid-plane where the 
vertical gravity vanishes. In other words, the MRI is an intrinsically 3-D 
instability, and it cannot operate efficiently in a 2-D magnetofluid except as 
artificial ``channel flows'' (see, e.g., Goodman \& Xu 1994) where the following 
considerations still apply.  

In circumstances where the flow is confined to 2-D, the turbulent resistivity 
$\eta$ is ``quenched,'' becoming proportional to its microscopic collisional 
value, although enhanced by a factor $(\delta B/B)^2$ when the fluctuations are 
large (Cattaneo \& Vainshtein 1991; Gruzinov \& Diamond 1994, 1966; Diamond, et 
al. 2005).  No matter how intricately turbulence distorts magnetic field lines 
in the remaining two (horizontal) directions,
electrically conducting particles cannot get off the field lines about which 
they gyrate, unless they are 
knocked off by microscopic physical collisions.  In 2-D MHD turbulence, an 
inverse cascade of squared magnetic potential exists alongside the familiar 
energy cascade to smaller scales.  This inverse cascade reflects a competition 
between the tendencies of velocity fluctuations to chop-up iso-contours of 
magnetic potential, thus producing smaller scales, and of magnetic loops to 
aggregate because of the attractive force between parallel lines of current, 
thus producing larger scales.   Thus, field lines never get chopped up 
systematically into small loops that can be reconnected much more quickly than 
the laminar dissipation of the mean fields.
In layered accretion, therefore, the transport of mass and angular momentum are 
subordinate to the diffusion of field, and the effective turbulent $\nu$ in the 
active surface layers may be constrained to be compatible with the microscopic 
collisional value of $\eta$. Because the entire layer is not involved in the 
relevant diffusive processes, a formulation that integrates through the vertical 
thickness cannot do justice to the real problem which has a bimodal vertical 
stratification. 

At a minimum, we should consider instead a two-layer description and introduce 
surface densities, 
$\Sigma_s$ and $\Sigma_m$, 
that describe respectively the columns through the upper and lower (active or 
live) surface layers and a middle (inactive or dead) layer, which sum to the 
total $\Sigma$ of the current formulation. In this picture, the values cited for the surface densities and magnetic-field strengths in \S 3.2 probably refer more to the active layer than they do to the total.  Although
the layer thickness expressed in terms of $\Sigma_s$ may be more-or-less fixed by the (external) sources
of ionization, the magnetic field strength $B_z$ is potentially adjustable as a function of $\varpi$ to give a constant accretion rate
$\dot M_{\rm d}$ in steady state, which yields an advantage of such a description of layered accretion over that given originally by Gammie (1996).

Our current calculations yield no constraint on the possible surface densiity of the inactive layer $\Sigma_m$.
Magnetic fields have freedom to
move with respect to the nearly un-ionized gas in $\Sigma_m$ that rotates
at near Keplerian speeds.  Shear instabilities  arising from upper and lower surfaces that
rotate slower than the midplane layers could lead to breaking radial buoyancy waves that
provide torques to redistribute the
angular momentum of the ``dead gas'' (Vishniac \& Diamond 1989).  The coupling 
provided by the excitation of waves in dead zones has been explicitly demonstrated in the 
simulations by Fleming \& Stone (2003) and W\"unsch et al.~(2006). In 
a more simplified approach that would not attempt to resolve the internal 
structure of the upper and lower layers, the layer with surface density 
$\Sigma_s$, would be described by the equations given in this paper, except again for a frictional term coupling them to $\Sigma_m$.  With 
$\Sigma_m$ included as a frictional load, the net effect would be as a variable 
$D$
in the single-layer description of the total surface density $\Sigma$. In other 
words, $\nu$ is slaved to $\eta$ in $\Sigma _s$ in the active, but geometrically 
thin, surface layers, and $\eta$ is given by its collisional value.  Then $D$ is 
simply a defined quantity in the current formula (\ref{def_nu}) for the 
relationship between $\nu$, $B_z$, $z_0$, $\Omega$, and {\it total} $\Sigma$. 
Such a two-layer model, with enough microphysics to specify the collisional 
value of $\eta$ in a complex, dusty, plasma, would allow us to calculate the 
variation of the effective $D$ with $\varpi$ in our current one-layer 
description.  

Figure 4 gives the estimate by Sano et al. (2000) of the microscopic collisional 
value of $\eta$ in the mid-plane of a Hayashi model solar nebula with dust of 
unagglomerated interstellar size and ionized by Galactic cosmic rays. The 
collisional resistivity in the inner disk ($\varpi < 3$ AU) is 
consistent with the magnitude $\sim 10^{20}$ cm$^2$ s$^{-1}$ cited by Shu et al. 
(2006) as needed for dynamic disk formation. The collisional resistivity beyond 
$\sim 10 - 20$ AU is lower than the values needed for the low-mass protostar 
model (long dashes with $D$ = 1) or for the T Tauri model (short dashes with $D 
= 10^{-2}$ and $D = 10^{-3}$).  In the interpretation of this section, then, 
once disk formation has occurred, its accretion resistivity would need to arise 
from MRI turbulence at large radii, whereas interior to $10-20$ AU, dead zones 
may be present and collisional resistivities may be adequate for the needed 
field diffusion even in the thin surface layers where viscous accretion is 
active.

An interesting question then arises as to what can initiate a transition between 
a low-state and a high-state of accretion.  It is natural to expect the 
transition to originate at a boundary between dead zones and live zones. By 
definition, such boundaries are not thin layers in any description where 
vertical stratification matters. The electric fields experienced by charged 
particles forced by collisions to rotate in the dead zone relative to the 
magnetic field, at speeds characterizing the large slip between the magnetically 
coupled active layer $\Sigma_s$ and the magnetically decoupled layer $\Sigma_m$, 
may generate suprathermal particles.  These suprathermal particles might produce 
the ionization that converts $\Sigma_m$ into a better conducting medium. 
Unfortunately, the existing numerical simulations of dead zones do not help us 
much in the latter regard because the acceleration of suprathermal particles 
requires a kinetic treatment, not just a (magneto)hydrodynamic one.  Moreover, 
the large slip is missing in the local simulations of Fleming \& 
Stone (2003); and the magnetic field is missing in the global simulations of 
W\"unsch et al.~(2006).  Heating by the resulting enhanced accretion may further 
enhance the development of the three-dimensional, turbulence of the type with 
which we started the discussion of this section. The boundary would then eat its 
way radially into the zones that were previously dead. An interesting 
theoretical goal would be to see how this transition between low states and high 
states works in detail and whether an FU Orionis outburst begins inside-out or 
outside-in since sufficiently-ionized regions from conventional sources exist on 
both sides of normal dead zones. 

\section{SUMMARY AND CONCLUSIONS}

The discussion of \S 4 represents our attempt to resolve the conflicts imposed 
by the following separate issues:

\noindent
{\it a}) the existence of a definite relationship between $\nu$ and $\eta$ in 
steady state,

\noindent
{\it b)} the evidence that the common diffusion coefficient $D$ has a value of 
order unity in some systems and much less than unity in others,

\noindent
{\it c)} the fact that MHD turbulence has a very different character in 2-D 
compared to 3-D,

\noindent
{\it d)} the suggestion that $\eta$ may be limited to have essentially its 
microscopic collisional value in layered accretion,

\noindent
{\it e)} the difficulty of weak magnetic coupling when ``dead zones'' arise in 
YSO disks.

\noindent
Our suggestions in \S 4 are therefore as much a roadmap of needed future 
research as they are a catalogue of the mysteries of the present and the past. 

In retrospect, the biggest mystery concerns the most observationally 
well-studied disks associated with star formation, those in T Tauri systems.  We 
may phrase the conundrum as follows. Diffusive processes cannot remove angular 
momentum or magnetic flux from the system; they can only redistribute them 
within the system.  In a closed system, T Tauri stars represent an end game for 
viscous resistive disks whose mass steadily drains into the central star, but 
whose magnetic flux and angular momentum, inherited by gravitational collapse 
from the interstellar medium, remain more-or-less trapped in the surrounding 
disk.  Such a situation must result eventually in a magnetically dominated disk.  
To prevent the residual disk from spreading to very large radii in a fixed 
amount of time, then demands inefficient diffusion (small $D$).  The 
astronomical challenge therefore becomes to explain
why there are two physical states of accretion, an active state (protostars, FU 
Orionis outbursts), characterized by a $D$ with a more ``natural'' value $\sim 
1$, and an inactive state (T Tauri stars), characterized by a $D\ll 1$. 

The conventional assessment is that ``dead zones'' provide the resolution for 
why the mass-accretion rate is so low in T Tauri disks observationally, or why 
the effective $D \ll 1$ in our language.  But if this is the correct answer, 
then why should $D$ ever be as large as unity in other contexts, the most 
obvious being FU Orionis outbursts? These disks have even higher column 
densities of disk matter that can shield external sources of ionization, 
principally, X-rays and cosmic-rays. Why aren't they even more full of dead 
zones?
We have proposed exploring the possibility that the high states of disk 
accretion correspond to the removal of such barriers.  Perhaps a fraction of the 
energy released in the resistive dissipation of stressed fields accelerates 
suprathermal particles and thus provide a level of {\it in situ}
ionization much in excess of the sources considered in conventional solar nebula 
models.   Thus, the MRI mechanism, properly generalized to include the 
dissipation of currents generated by the stressing of mean fields from viscous 
accretion, may contain its own solution to the challenge posed by low ionization 
(see also, Fromang et al.~2002). Heating from enhanced accretion may also help 
with the ionization of trace species such as lithium and potassium.  Indeed, 
since any bootstrap mechanism allows the potential of a runaway -- more 
ionization $\rightarrow$ more coupling $\rightarrow$ more ionization, etc. -- 
this proposal also offers an opportunity, perhaps, to understand why disk 
accretion in YSOs can alternate between low and high states of accretion, 
characterized generically by T Tauri stars and FU Orionis outbursts.

The complementary difficulties noted above are related to a serious problem 
noticed by King, Pringle, \& Livio (2007). The effective  $\alpha_{\rm ss}$ in a 
Shakura-Sunyaev prescription for disk viscosity has to be of order 0.1 to 0.4 to 
explain the empirical facts known about thin, fully-ionized, accretion disks in 
many astronomical contexts, yet the equivalent alpha from almost all MRI 
simulations to date is lower typically by one or more orders of magnitude (e.g., 
as summarized by Gammie \& Johnson 2005 and modeled by Fromang et al.~2002).  By 
coincidence, low values for $\alpha_{\rm ss} \sim 10^{-2}$ are empirically 
acceptable for modeling T Tauri disks (see, e.g., Hartmann et al.~1998), because 
such disks have extensive dead zones, in which only a small fraction of the 
total surface density is actively accreting.  But if the estimates from MRI 
simulations were applied only to the active layers, the net effective value of 
$\alpha_{\rm ss}$ (in the sense of a combined two-layer model) would have been more like $10^{-4}$ than $10^{-2}$.  

Among the possible resolutions mentioned by King et al. (2007) are stronger 
magnetic fields and global rather than local simulations for the MRI that occurs 
in realistic systems.  The results of the present paper (see especially the 
discussion of \S\S 2.1 and 4.1) strongly support such a resolution of the 
existing paradox, at least for the field of star formation.  Most MRI 
simulations ignore the presence of a nonzero magnetic flux that threads through 
the disk, carried in by the process of gravitational collapse. As demonstrated 
in this paper, the presence of an externally supplied magnetic field makes the 
self-consistent dynamics considerably more subtle than the simplest notion of 
the MRI extant in the literature. In particular, the accretion flow generates, 
on either side of the mid-plane, a mean radial field $B_\varpi$ from the mean 
vertical field $B_z$ because of
the inward drift and the imperfect tendency toward field freezing. This mean 
radial field, whose surface value
is denoted by the symbol $B_\varpi^+$ in this paper and whose properties can be 
deduced only by a global
calculation that takes proper consideration of the vacuum conditions above (and 
below) the plane
of the disk, sets the scale for turbulent fluctuations (if MRI arises) and has 
three important consequences. 

First, as emphasized in \S 4.1, the resultant poloidal-field configuration can 
spawn magnetic loops, possessing a radial component
of the magnetic field $\delta B_\varpi$ whose amplitude is proportional to 
$B_\varpi^+$.  The loop
can be stretched in the azimuthal direction
by the differential rotation in the disk to produce an azimuthal field $\delta 
B_\varphi$ that has a
systematic orientation with respect to $\delta B_\varpi$.  The correlation of 
$\delta B_\varpi$ and $\delta B_\varphi$
then exerts a Maxwell stress much larger than the corresponding values obtained 
in simulations of MRI where there is
no external field $B_z$ to set a scale for $B_\varpi^+$.  The Maxwell stress 
leads to angular momentum transport that yields the original accretion 
responsible for the generation of mean $B_\varpi^+$ from mean $B_z$.

Second, the resultant poloidal-field configuration introduces current flows that 
can be dissipated by resistive effects.
An important finding of our study, extending the work of LPP, is that the ratio 
of the effective resistivity, $\eta$, to the turbulent viscosity, $\nu$, must 
have a well-specified value in steady state that depends on the local aspect 
ratio of the disk (vertical thickness to radius). The exploration of the 
implications of this result for the turbulent microphysics of the problem, in 
particular,
how the microscopic dynamics of the current loops can automatically adjust to 
the requirements of the macroscopic problem, needs further theoretical study, 
best supplemented by numerical simulation.

Third, the resultant poloidal-field configuration produces a change in the
radial force balance, giving a deviation from the traditionally
invoked Keplerian profile. This deviation is not easily detectable 
observationally because the resulting
rotation law has in steady state the same power-law dependence with radius as a 
true Kepler law, but the
coefficient is smaller.  Thus, even if it were present, observers would tend to 
attribute the result
to the mass of the central object being smaller than its actual value, or to the 
disk being inclined by
a lesser amount than in reality.  Nevertheless, it would be illuminating to find 
such
an effect in the YSO disks that have the smallest mass-to-flux ratios.  Indeed, 
a deduction of sub-Keplerian rotation of the disk
may already have been made in the case of FU Orionis (Donati et al. 2005), but 
the proper interpretation of the phenomenon in this case may be complicated by 
the interaction of the disk field with the imperfectly squashed stellar fields 
of the central object.
Finally, it has not escaped our attention that significant departures from true 
Keplerian rotation of a YSO disk
may have important consequences for the problems of binary-star and 
planetary-system formation and evolution,
particularly with regard to the difficult issues of orbit migration and 
eccentricity pumping (e.g., Goldreich \& Sari 2004).
The inward drift of solids from sub-Keplerian regions into the dead zones of the 
problem, which rotate more nearly at Keplerian speeds, may give an additional 
reason to focus on the importance of dead zones for the problems of planetesimal 
and planet formation (e.g., Youdin \& Shu 2002, Pudritz \& Matsumura 2004).

\bigskip\noindent
We thank Steve Lubow for pointing us to the paper by Ogilvie (1997).  FS is 
grateful for the support of the UCSD physics department; DG and SL wish to thank 
the Astronomy Department of the University of California at Berkeley for 
hospitality. SL also acknowledges support from CONACyT 48901 and PAPIIT-UNAM 
IN106107; AG, from NSF Grant AST-0507423; and PD, from US Department of Energy 
Grant No. FG02-04ER 54738 . 

\appendix  

\section{ALTERNATIVE FORMULATIONS FOR TURBULENT DIFFUSION}

A turbulent, magnetized medium may not behave in 
the assumed model fashion,  with
diffusive fluxes proportional to a scalar diffusivity times the rate of spatial 
variation of the mean quantity that is being spread (Fick's law).  For example, 
Pessah et al. (2007) perform a quasi-linear analysis of the MRI with third-order 
closure in a simple shearing-box geometry.  They claim that the turbulent 
viscous stress $\Pi_{\varpi\varphi}$ depends on the rate of shear by a non-Fickian power $p$ 
that is different from 1.  If we denote $(\varpi /\Omega)\partial \Omega 
/\partial \varpi$ by $-S$, then their model yields $\Pi_{\varpi\varphi} = -\hat \nu
\Sigma \Omega S^p$ with $p$ between 3 and 4. However, {\it in a field of 
quasi-Keplerian rotation} where $\Omega \propto \varpi^{-3/2}$, $S$ is simply 
the number 3/2.  Thus, the Pessah et al. formalism satisfies, in practice, the 
usual Newtonian relationship, $\Pi_{\varpi\varphi} = \nu \Sigma \varpi \partial 
\Omega/\partial \varpi$, where $\nu = (3/2)^{p-1}\hat \nu$.  This simple 
transformation allows the translation of all of our results into the language of 
Pessah et al.

A similar remark applies to the induction equation (\ref{indeq}).  
Instead of the diffusion of vertical field, some mean-field formulations of turbulent MHD (see \S 4.1) envisage
the diffusion of tangential  current (per unit length).  From Ampere's law, $J_\varphi = 
(c/2\pi)B_\varpi^+$, and the diffusion of mean current is equivalent to the 
diffusion of the mean radial field.  Then, instead of equation (\ref{indeq}), we 
might postulate a diffusion equation of the form
\be
{\partial B_\varpi^+\over \partial t}+{1\over \varpi}{\partial \over \partial 
\varpi}\left( \varpi B_\varpi^+ u\right) =
{1\over \varpi}{\partial \over \partial \varpi} \left( \eta_J \varpi {\partial 
B_\varpi^+ \over \partial \varpi}\right),
\label{currentdiff}
\ee
where $\eta_J$ is the diffusivity for the turbulent diffusion of current.  In 
steady-state, radial force balance in a field of 
quasi-Keplerian rotation will still
require $B_\varpi^+ = I_\ell B_z \propto \varpi^{-(5+2n)/4}$ (see \S\S 1.3 and 
2).  In this case, the above equation is equivalent to equation (\ref{indeq}) if 
we identify
\be
\eta_J = {4I_\ell \over (5+2n)} \left({\eta \over z_0/\varpi}\right) = \left( 
{6\over 5+2n}\right)\nu,
\ee
where we have used equation (\ref{Prandtl}).
For $n =1/4$, the relationship between $\eta_J$ and $\nu$ is then $\eta_J = 
1.09\, \nu$, a result that we might call the ``Prandtl hypothesis'' because it 
differs only slightly from the naive guess that diffusion processes in a 
turbulent medium have equal steady-state diffusivities (cf.~the discussion in 
LPP). Thus, what seems more relevant than the specific turbulent diffusivities are the turbulent fluxes,
and how those fluxes relate to the spatial derivatives of mean-flow quantities.

\section{THE VALUE OF $D$ FROM MRI SIMULATIONS}

\begin{table}
\bc
\begin{tabular}{lcccc}
\multicolumn{5}{c}{Table 3. Estimates of $D$} \\
\hline
\hline
Run & $\beta(0)$ & $\langle\langle B_x B_y\rangle\rangle/\langle\langle 
B^2\rangle\rangle$ 
        & $\langle\langle B^2\rangle\rangle/\langle\langle B_z^2\rangle\rangle$ 
& $D$   \\ 
\hline
SHGB IZ1 & 100   & 0.145           & 26.7    & 0.2       \\
SHGB IZ3 &  25   & 0.139           &         &       \\
MS ZN2  & 100   & 0.0716/0.0958   &     &           \\
MS ZN1  & 25    & 0.0111/0.00586  & 40/26   & 0.03/0.01 \\
\hline
\label{table_mri}
\end{tabular}
\ec
\end{table}
 
In Table 3, we list the ratio of magnetic stress 
to magnetic energy density from Table 1 of Stone et al.~(1996; SHGB) 
and from Table 1 of Miller \& Stone~(2000; MS). Double angle-brackets 
indicate time and volume averages (mostly over two scale heights) 
for the two smallest values of the initial mid-plane beta parameter, 
$\beta(0)$, in these papers. SHGB use periodic boundary 
conditions in all three directions and a vertical box height three times 
the initial pressure scale height $H$. MS employ an  outgoing boundary condition at the top 
of a box of height $5H$.  In the Table, two averages are included for the results of MS, which apply to the 
regions $|z| < 2H$ (fourth row) and $|z| > 2H$ (fifth row).  Empty entries occur when the required 
information is not given in the original papers.

In the simulations, $B_x$ and $B_y$ contain
only fluctuating components, whereas $B_z$ and total $B$ contain both mean and 
fluctuating components, with the mean components being systematically destroyed by numerical 
reconnection as the simulations proceed because the starting distribution of $B_z$ alternates in sign radially.
If we equate the tangential Maxwell stress per unit circumferential length 
(which dominates the corresponding expressions for the Reynolds stress) to an equivalent viscous 
stress in the usual manner, we
obtain the following expression for $\nu$ 
for the case of quasi-Keplerian rotation,
\be
\nu = \frac{z_0}{3\pi\Sigma \Omega}
\langle\langle B_xB_y\rangle\rangle
\ee
Introducing a factor of $\langle \langle B^2\rangle \rangle$ on top 
and bottom to express the relevant ratios in the form of Table 3, we can now identify the coefficient $D$ in equation (\ref{def_nu}) as
\be
D = \frac{1}{6\pi}\frac{\langle\langle B_xB_y\rangle\rangle}{\langle\langle 
B^2\rangle\rangle} \, 
\frac{\langle\langle B^2\rangle\rangle}{\langle\langle B_z^2\rangle\rangle}.
\ee

The results in Table 3 give values of $D$ that differ by 
an order of magnitude, indicating that the MRI simulations are sensitive to the 
assumed boundary
conditions and to the size of the computational box.  As a formal result, the last column of Table 3 seems to suggest that $D$ is, at best, 0.2. However,
as emphasized in the text, simulations with net magnetic flux equal to zero do 
not correspond to the situation of interest for
our study. 

\section{VERTICAL STRUCTURE OF STRONGLY MAGNETIZED DISKS}

Inside a thin disk, the condition of vertical hydrostatic equilibrium reads
\be
{\partial\over \partial z}\left( P+{B_\varpi^2+B_z^2\over 8\pi}\right) = 
-{GM_*z\rho\over \varpi^3},
\label{hydrostatic}
\ee
where $P$ is the thermal gas pressure and $B_\varpi$ is a function of $z$ equal 
to 0 at $z=0$ and to $B_\varpi^+$
at the disk's surface, whereas $B_z$ may be taken to be a constant over the same 
range of $z$. If we integrate equation
(\ref{hydrostatic}) from $z=0$ to the surface under the boundary conditions that 
$P(0) = a^2\Sigma/2z_0$ (which defines what we mean by $a^2$) and $P = 0$ at the 
disk's surface (which defines what we mean by the surface), we get
\be
{I_\ell^2B_z^2\over 8\pi} - {a^2\Sigma 
\over 2z_0} = -{GM_*\Sigma z_0\over 4\varpi^3},
\label{vertbal}
\ee
where we have defined $z_0$ by requiring the integral of $z\rho$ from $z=0$ to 
the surface of the disk yield
$(z_0/2)(\Sigma/2)$.  With $B_\varpi^+ = I_\ell B_z$, the equation of radial force 
balance (\ref{centrifugal}) reads
\be
{I_\ell B_z^2\over 2\pi \Sigma} = (1-f^2){GM_*\over \varpi^2}.
\label{radforcebal}
\ee
Elimination of $B_z^2$ from equations (\ref{vertbal}) and (\ref{radforcebal}), with $z_0 = A\varpi$, then gives equation 
(\ref{implicit_a2}).

\newpage
\begin{figure}
\plotone{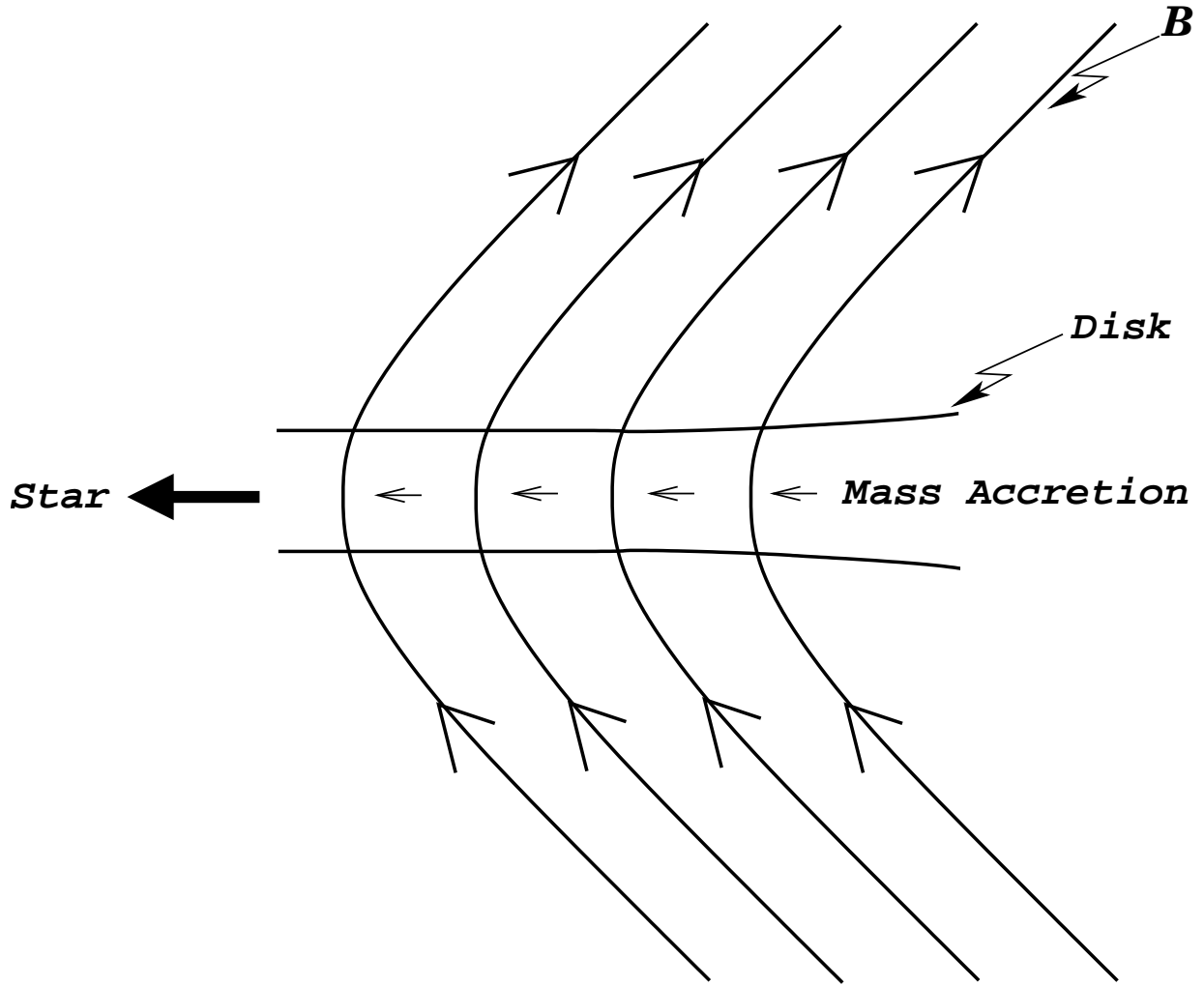}
\caption{
Schematic diagram of accretion flow in a disk threaded by
magnetic flux accumulated by the process of star formation.}
\label{fig1}
\end{figure}

\begin{figure}
\plotone{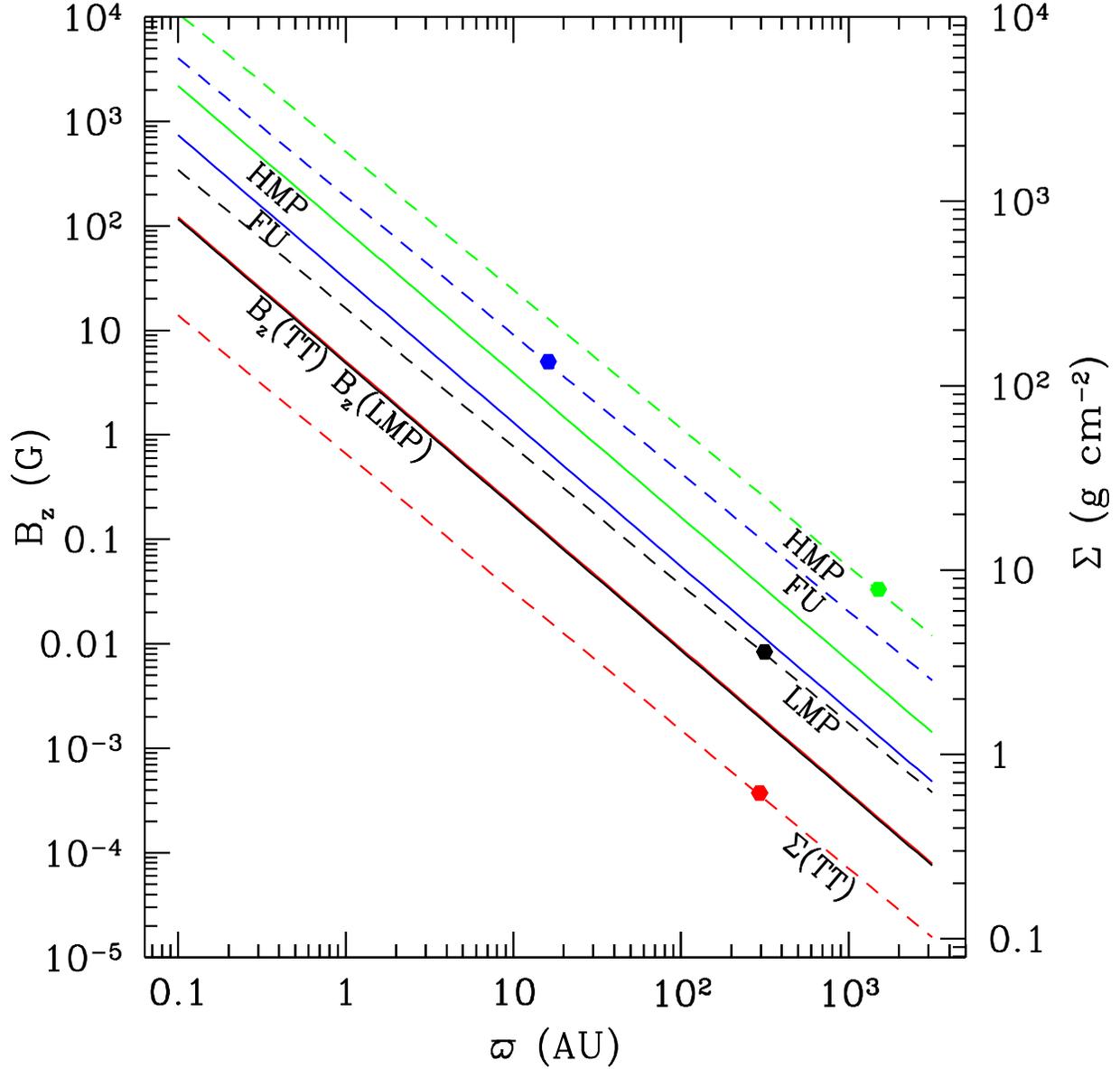}
\caption{The vertical component of the magnetic
field $B_z$ (solid curves) and the surface density
$\Sigma$ (dashed curves) plotted against the radius $\varpi$ in the
steady-state disks of the four models of Table 2.  The hexagons mark
the location where $R_\nu = R_\Phi$ in the T Tauri (red curves),
low-mass protostar (black curves), FU Orionis (blue curves), and
high-mass protostar (green curves) models. The slopes are $-11/8$ and
$-3/4$, respectively, for $\log B_z$ and $\log \Sigma$ versus $\log \varpi$.}
\label{fig2}
\end{figure}

\begin{figure}
\plotone{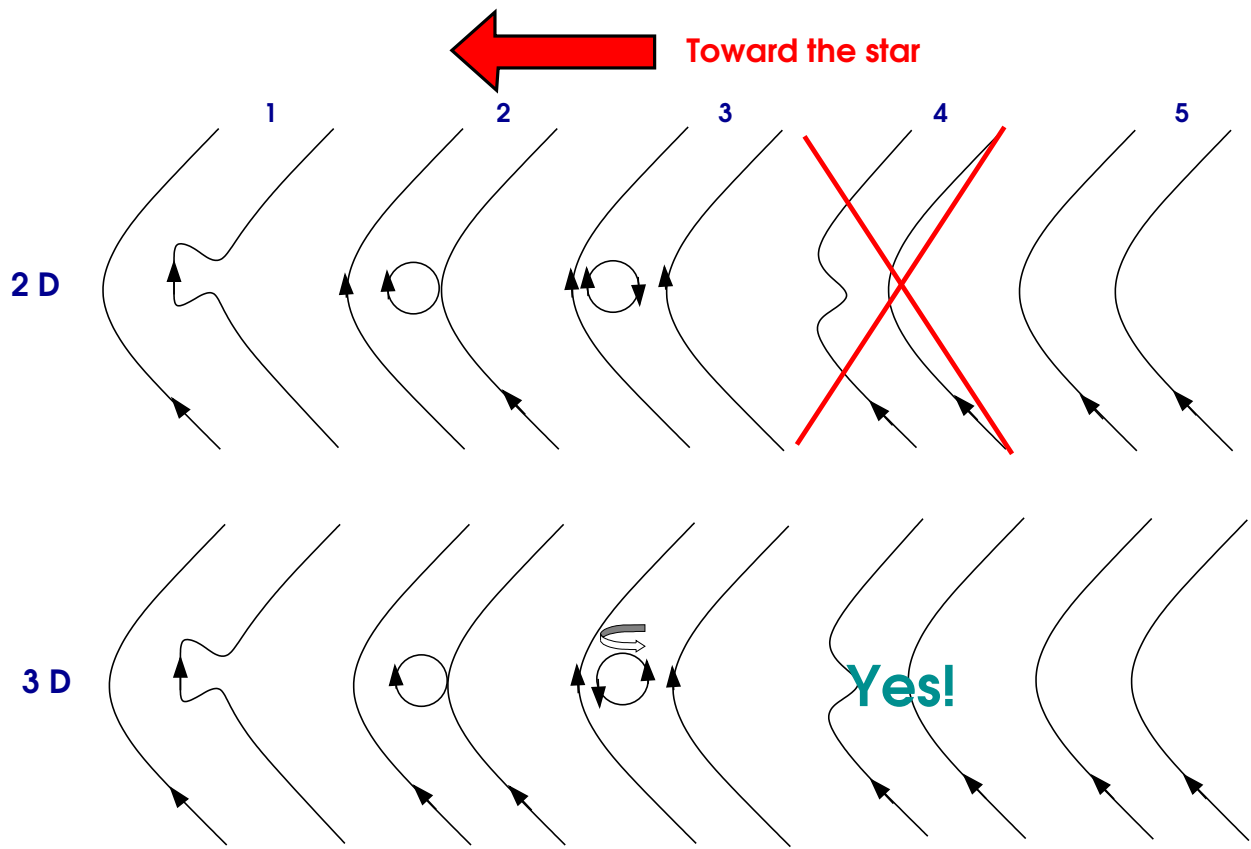}
\caption{Schematic diagram of scenarios by which field loops are
created by magnetohydrodynamic turbulence when the mean field is
strong: (top) in 2-D by stretch, pinch, disconnect and (bottom) in 3-D
by stretch, pinch, disconnect and twist, reconnect, relax.  The
depiction is the meridional plane $(\varpi, z)$, except for the twist
indicated by the block arrow, which occurs out of the plane of
the paper in the $\varphi$ direction because of differential rotation.
Notice the bias for forming the
loop on the side closer to the star because of the accretion flow. This
bias causes the diffusive flux to flow in the correct direction
relative to the curl of mean ${\bf B}$. Because the loop in the top
diagram does not experience the twist operation, it has the wrong
orientation to reconnect with the mean field downstream of the mean
accretion flow since the fields point up on both sides of the target
contact point. The twist in the bottom diagram gets the fields oriented
in opposite directions at the target reconnection point, which results in the
green ``yes'' sign to proceed to steps 4 and 5.}
\label{fig3}
\end{figure}

\begin{figure}
\plotone{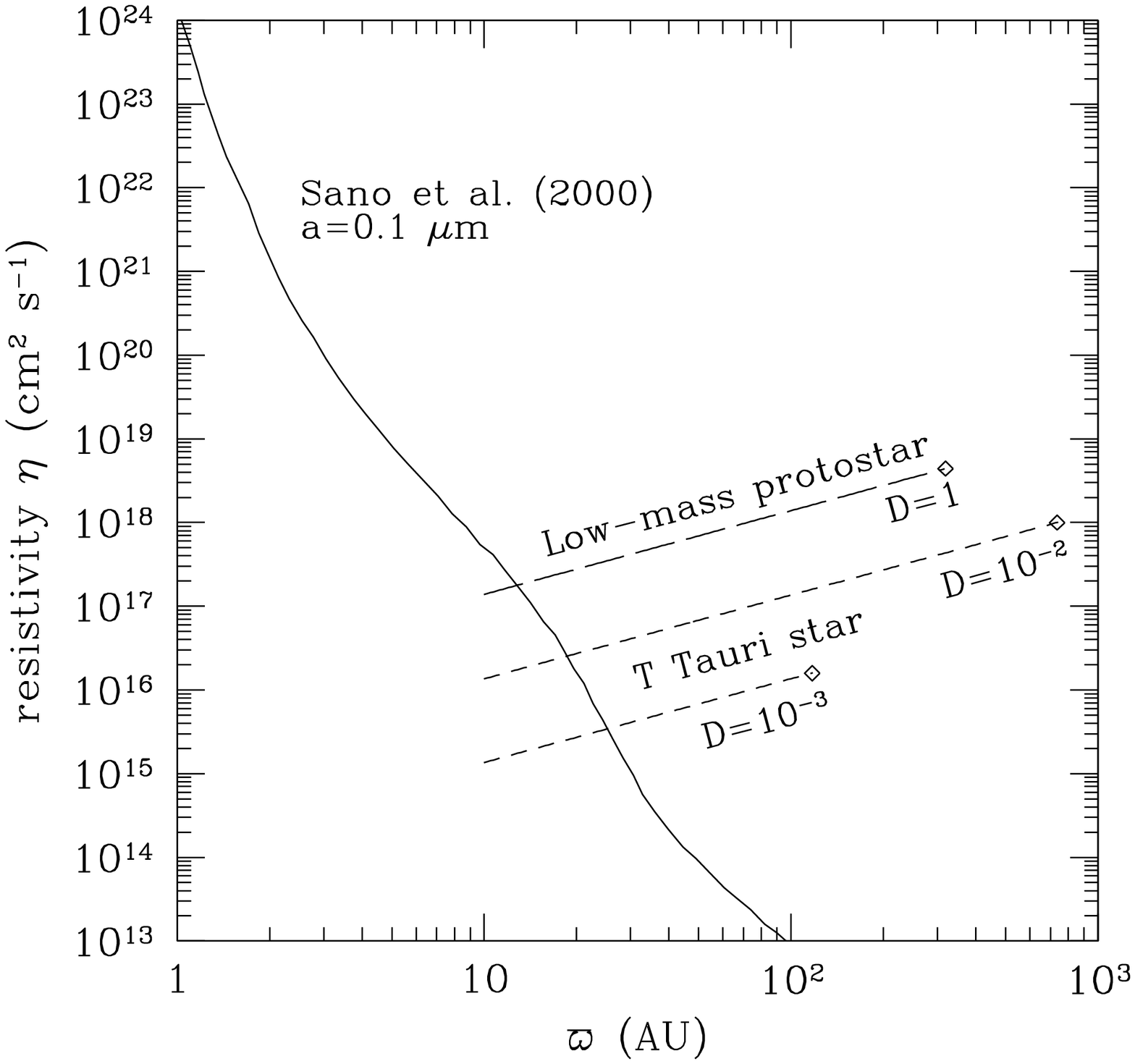}
\caption{Comparison of collisional resistivity (solid curve) applicable
to the Sano et al. (2000) model of the minimum solar nebula, which
includes the effects of cosmic grains of typical interstellar size ($a
= 0.1 \, \mu{\rm m}$) with the required turbulent values in the models of
\S 3 of a low-mass protostar (long dashes) with $D = 1$ and a T Tauri
star (short dashes) with $D = 10^{-2}$ (upper curve) and $D = 10^{-3}$
(lower curve).  The hexagons mark the corresponding locations of $R_\Phi$
where the disk holds the trapped flux corresponding to a dimensionless 
mass-to-flux ratio
$\lambda_0$ = 4.}
\label{fig4}
\end{figure}

\begin{thebibliography}{}

\bibitem{}
Adams, F.~C., Proszkow, E.~M., Fatuzzo, M., Myers, P.~C. 2006, ApJ, 641, 504 

\bibitem{}
Andrews, S. M.,  Williams, J. P. 2007, Ap J, 659, 705

\bibitem{}
Bachiller, R. 1995, ARAA, 34, 111

\bibitem{bh98}
Balbus, S. A., Hawley, J. F. 1998, Rev. Mod. Phys., 70, 1

\bibitem{}
Basu, S., Mouschovias, T. Ch., 1994, ApJ, 432, 720

\bibitem{}
Binney, J., Tremaine, S. 1987, Galactic Dynamics, (Princeton University Press), 
pp. 120-126

\bibitem{bp82}
Blandford, R. D., Payne, D. G. 1982, MNRAS, 199, 883

\bibitem{cv91}
Cattaneo, F., Vainshtein, S.~I. 1991, ApJ, 376, 121

\bibitem{}
Chan, K. L.,  Henriksen, R. N. 1980, ApJ. 241, 534

\bibitem{}
Cisowski, S. M., Hood, L. L. 1991, in The Sun in Time, ed. C. P. Sonett, M. S. 
Giampapa, M. S. Matthews
(University of Arizona Press), p. 761 

\bibitem{}
D'Alessio, P., Calvet, N., Hartmann, L., Lizano, S., Canto, J. 1999, 527, 893

\bibitem{dal05}
Diamond, P.H., Malkov, M., 2003, Phys. Plasmas 10, 2322

\bibitem{dhk05}
Diamond, P.~H., Hughes, D.~W., Kim, E. 2005 in Fluid Dynamics and Dynamics in 
Astrophysics and Geophysics, (Soward, A., Jones, C., Hughes, D., Weiss, N., 
eds.) 145

\bibitem{d05}
Donati, J. F., Paletou, F., Bouvier, J., Ferreira, J. 2005, Nature, 438, 466

\bibitem{}
Dutrey, A., Guilloteau, S.,  Ho, P.~2007, in Protostars \& Planets V, eds. B.~Reipurth, D.~Jewitt, \& K.~Keil
(Tucson, Univ. Arizona Press), p. 951

\bibitem{}
Edris, K. A., Fuller, G. A.,  Cohen, R. J. 2007, A\&A, 46, 865

\bibitem{}
Evans, N. J. 1999, ARAA, 37, 311

\bibitem{}
Ferreira, J., Dougados, C.,  Cabrit, S. 2006, A\&A, 453, 785

\bibitem{fsh00}
Fleming, T.~P., Stone, J.~M., Hawley, J.~F. 2000, ApJ, 530, 464 

\bibitem{}
Fleming, T.~P.,  Stone, J.~M. 2003, ApJ, 585, 908

\bibitem{}
Font, A. S., McCarthy, I. G., Johnstone, D., Ballantyne, D. R. 2004, 607, 890 

\bibitem{}
Fromang, S., Terquem, C., Balbus, S. A. 2002, MNRAS, 329, 18

\bibitem{}
Fromang, S., Terquem, C, Nelson, R. P. 2005, MNRAS, 363, 943 

\bibitem{g06}
Galli, D., Lizano, S., Shu, F. H., Allen, A. 2006, ApJ, 647, 374

\bibitem{}
Gammie, C. F. 1996, ApJ, 457, 355

\bibitem{}
Gammie, C. F., Johnson, B. M., 2005, ASP, 341, 145

\bibitem{g06}
Girart, J. M., Rao, R., Marrone, D. 2006, Science, 313, 812

\bibitem{}
Goldreich, P. Lithwick, Y., Sari, R. 2004, ARAA, 42, 549

\bibitem{}
Goldreich, P., Lynden-Bell, D. 1969, ApJ, 156, 59

\bibitem{}
Goldreich, P., Sari, R. 2004, ApJ, 606, L77

%\bibitem{}
%Goldreich, P., Sridhar, S. 1995, ApJ, 438, 763

\bibitem{}
Goldreich, P., Sridhar, S. 1997, ApJ, 485, 680

\bibitem{}
Goodman, J.,  Xu, G. 1994, ApJ, 432, 213

\bibitem{gbw99}
Goodson, A. P., Bohm, K. H., Winglee, R. M. 1999, ApJ, 524, 142

\bibitem{gd94}
Gruzinov, A.V., Diamond, P.H., 1994, Phys. Rev. Lett. 72, 1651

\bibitem{gd96}
Gruzinov, A.~V., Diamond, P.~H. 1996, Phys. Plasmas 3, 1853

\bibitem{}
Gullbring,  E., Hartmann, L., Briceno, C., Calvet, N. 1998, ApJ, 492, 323

\bibitem{}
Haisch, K. E., Lada, E. A., Lada, C. J. 2001, ApJ, 553, L153

\bibitem{}
Halwachs, J. L., Arenou, F., Mayor, M., Udry, S., Queloz, D. 2000, A\&A, 355, 
581

\bibitem{}
Hartmann, L., Calvet, N., Gullbring, E., D'Alessio, P.\ 1998, ApJ 495, 385 

\bibitem{hk96}
Hartmann, L., Kenyon, S. J. 1996, ARAA, 34, 207

\bibitem{hb91}
Hawley, J. F., Balbus, S. A. 1991, ApJ, 381, 496

\bibitem{hnn85}
Hayashi, C., Nakazawa, K., Nakagawa, Y. 1985, 
in Protostars \& Planets II, eds. D.~C.~Black, M.~S.~Matthews 
(Tucson: Univ. Arizona Press), 1100 

\bibitem{}
Herbig, G. H. 1977, 217, 693

\bibitem{}
Herbig, G. H., Petrov, P. P., Duemmler, R. 2003, ApJ, 595, 284

%\bibitem{hc99}
%Hutawarakorn, B., Cohen, R. J. 1999, MNRAS, 303, 845

\bibitem{hc05}
Hutawarakorn, B., Cohen, R.~J. 2005, MNRAS, 357, 338 

\bibitem{}
Igea, J.,  Glassgold, A. E. 1999, ApJ, 518, 848

\bibitem{}
Jijina, J., Myers, P. C.,  Adams, F. C. 1999, ApJS, 125, 161

%\bibitem{}
%Johansen, Klahr,  Mee 2006, MNRAS, 370, L71

\bibitem{}
Jorgensen, J. K., Bourke, T. L., Myers, P. C., Di Francesco, J., van Dishoeck, 
E. F., Lee, C. F., Ohashi, N., Schoier, F. L., Takakuwa, S., Wilner, D. J., 
Zhang, Q. 2007, Astro-ph 070115

\bibitem{}
King, A. R., Pringle, J. E., Livio, M. 2007, Astro-phy 0701803v1

\bibitem{kp00}
K\"onigl, A., Pudritz, R. E. 2000, 
in Protostars \& Planets IV, eds. V.Mannings, A. P. Boss, S. S. Russell (Univ. 
Arizona Press), p. 759

\bibitem{kk02}
Krasnopolsky, R., K\"onigl, A. 2002, ApJ, 580, 987

\bibitem{}
Krasnopolsky, R., Gammie, C. F. 2005, 635, 1126

\bibitem{}
K\"uker, M., Henning, Th.,  R\"udiger G. 2004, Ap\&SS, 292, 599

\bibitem{k06}
Kwan, J., Fischer, W., Edwards, S., Hillenbrand, L. 2006, 
in Protostars \& Planets V, ed. B. Reipurth (Tucson: Univ. Arizona Press), in 
press.

\bibitem{l06}
Lada, C. J. 2006, ApJ, 640, L63

\bibitem{}
Levy, E. H., Sonett, C. P., in Protostars \& Planets, ed. T. Gehrels (Tucson: 
University of Arizona Press), p. 516

\bibitem{}
Lin, D. N. C. 2006, in Planet Formation. eds. H. Klahr, W. Brandner (Cambridge 
University Press), p. 256

\bibitem{}
Lin, D. N. C., Papaloizou, J. 1995, ARAA, 33, 505

\bibitem{}
Lin, D. N. C., Papaloizou, J. 1996, ARAA, 34, 703

\bibitem{}
Lissauer, J. J. 1993, ARAA, 31, 129

\bibitem{lrl07}
Long, M., Romanova, M. M., Lovelace, R. V. E. 2005, ApJ, 634, 1214
  
\bibitem{}
Lubow, S. H., Papaloizou, J. C. B.,  Pringle, J. E. 1994, MNRAS, 267, 235 
(LPP)

\bibitem{lb69}
Lynden-Bell, D. 1969, Nature, 223, 690

\bibitem{lbp74}
Lynden-Bell, D., Pringle, J. E. 1974, MNRAS, 168, 603

\bibitem{}
Marcy, G. W.,  Butler, P. R. 2000, PASP, 112, 768

\bibitem{}
McKeegan, K. 2006, AGU meeting, abstract P52B-03

\bibitem{ms56}
Mestel, L., Spitzer, L.1956, MNRAS, 116, 503

\bibitem{ms00}
Miller, K. A., Stone, J. M. 2000, ApJ, 534, 398 (MS)

\bibitem{n96}
Najita, J., Carr, J., Glassgold, A. E., Shu. F. H., Tokunaga, A.T. 1996, ApJ, 
462, 919

\bibitem{}
Nakano, T., Nakamura, T. 1978, PASJ, 31, 697

\bibitem{}
Ogilvie, G. I. 1997, MNRAS, 288, 63

\bibitem{}
Osorio, M., Lizano, S., D'Alessio, P. 1999, ApJ, 525, 808

\bibitem{}
Parker, E. N. 1955, 122, 293

\bibitem{}
Parker, E. N. 1963, Interplanetary Dynamical Processes (New York: Interscience 
Pub)

\bibitem{}
Parker, E. N. 1966, ApJ, 145, 811

\bibitem{pcp07}
Pessah, M. E., Chan, C., Psaltis, D. 2007, astro-ph/0612404

\bibitem{}
Popham,  R., Kenyon, S., Hartmann, L., Narayan, R. 1996, ApJ, 422

\bibitem{}
Pudritz, R. E.,  Matsumura, S. 2004, in Gravitational Collapse: From Massive 
Stars to Planets, eds. G. Garcia-Segura, G. Tenario-Tagle, J. Franco, \& H. W. 
Yorke, Rev. Mex. A. A. (Conf. Series), 22, 108 

\bibitem{p25}
Prandtl, L. 1925, Z. angew. Math. Mech., 5, 136 

\bibitem{}
Pringle, J. 1981, ARAA, 19, 137

\bibitem{}
Rafikov, R. 2007, ApJ, submitted

\bibitem{}
Reipurth, B., Bally, J. 2001, ARAA, 39, 403

\bibitem{}
Rudiger, G., Shalybkov, D. A. 2002, A\&A, 393, L81

\bibitem{s00} 
Sano, T., Miyama, S.~M., Umebayashi, T., Nakano, T. 2000, ApJ, 543, 486 (S00)

\bibitem{ss73}
Shakura, N. I., Sunyaev, R. A. 1973, A\&A, 24, 337

\bibitem{ssg98}
Shang, H., Shu, F. H., Glassgold, A. E. 1998, ApJ, 439, 91

\bibitem{}
Shu, F. H. 1995, Rev Mex AA, 1, 375

\bibitem{}
Shu, F. H., Adams, F. C., Lizano, S. 1987, ARAA, 25, 23

\bibitem{s99}
Shu, F. H., Allen, A., Shang, H., Ostriker, E. C., Li, Z. Y. 1999, 
in The Origin of Stars and Planetary Systems, eds. C. Lada, N. Kylafis (Kluwer), p. 193.

\bibitem{s06}
Shu, F. H., Galli, D., Lizano, S., Cai, M. 2006, ApJ, 647, 382

\bibitem{sl97}
Shu, F. H., Li, Z.-Y. 1997, ApJ, 475, 251

\bibitem{sla04}
Shu, F. H., Li Z.-Y., Allen, A. 2004, ApJ, 601, 930

\bibitem{}
Shu, F., Najita, J., Ostriker, E., Wilkin, F., Ruden, S., Lizano, S. 1994, ApJ, 
429, 781

\bibitem{}
Shu, F. H., Shang, H., Glassgold, A. E., Lee, T. 1997, Science, 277, 1475

\bibitem{}
Shu, F. H., Shang, H., Lee, T. 1996, Science, 271,1545

\bibitem{}
Shu, F. H., Tremaine, S., Adams, F. C., Ruden, S. P. 1990, ApJ, 358, 495

\bibitem{}
Silver, L. J.,  Balbus, S. A. 2006, in Proc. Ann. Meeting French Soc. A\&A, 
eds. D. Barret, F. Casoli, G. Lagache, A. Lecavelier, \& L. Pagani, p. 107

\bibitem{}
Stacey, F. D. 1976, AREPS, 4, 147

\bibitem{}
Stauber, P., Doty, S.~D., van Dishoeck, E.~F., Jorgensen, J.~K., Benz, A.~O. 
2007, A\&A, in press

\bibitem{SHGB}
Stone, J. M., Hawley, J. F., Gammie, C. F., Balbus, S. A. 1996, ApJ, 463, 656 
(S96)

\bibitem{Strauss}
Strauss, H.R., 1976, Phys. Fluids 19, 134

\bibitem{}
Terquem, C. 2003, MNRAS, 341, 1157

\bibitem{}
Toomre, A. 1964, ApJ, 139, 1217

\bibitem{u06}
Ustyugova, G.~V., Koldoba, A.~V., Romanova, M.~M., Lovelace, R.~V.~E.\ 2006, 
ApJ, 646, 304 

\bibitem{}
van Ballegooijen, A.~A. 1989, in Magnetic Fields in Astrophysics, ed. G.~Belvedere (Kluwer), p. 99

\bibitem{vd}
Vishniac, E.T., Diamond, P.H., 1989, Ap.J. 347, 435

\bibitem{}
Vorobyov, E. I.,  Basu, S. 2006, ApJ, 650, 956

\bibitem{wn99}
Wardle, M., Ng, C. 1999, MNRAS, 303, 239

\bibitem{}
Wardle, M., K\"onigl, A. 1993, ApJ, 410. 218

\bibitem{}
White, R. J., Greene, T. P., Doppmann, G. W., Covey, K. R., Hillenbrand, L. A. 
2007, Protostars \& Planets V, eds. B. Reipurth, D. Jewitt, \& K. Keil (Tucson: 
University of Arizona Press), p.117

\bibitem{}
W\"unsch, R., Gawryzczak, A., Klahr, H.,  R\'ozyczka, M. 2006, MNRAS, 367, 773

\bibitem{}
Youdin, A.,  Shu, F. H. 2002, ApJ, 580, 494

\bibitem{}
Young, C. H., Evans, N. J. 2005, 627, 293

\bibitem{}
Young, C. H., Shirley, Y. L, Evans, N. J., Rawlings, N. M. C. 2003, ApJS, 145, 
111

\bibitem{}
Zolensky, M. E. et al. 2006, Science, 314, 1735


\end{thebibliography}
\end{document}